\documentclass[12pt,preprint]{aastex}

\shorttitle{Stable planetary geometry in 2:1 mean motion
resonance}

\shortauthors{Ji Jianghui et al.}

\begin{document}

\title{The Stable Planetary Geometry of the Exosystems in 2:1 Mean Motion Resonance}
\author{JI  Jianghui\altaffilmark{1,2}, Kinoshita Hiroshi\altaffilmark{3},
LIU Lin\altaffilmark{4,2}, Nakai Hiroshi\altaffilmark{3}, LI
Guangyu\altaffilmark{1,2}}

\altaffiltext{1}{Purple  Mountain Observatory, Chinese  Academy of
Sciences, Nanjing 210008, China;jijh@pmo.ac.cn}

\altaffiltext{2}{National Astronomical Observatory, Chinese
Academy of Sciences, Beijing 100012, China}

\altaffiltext{3}{National Astronomical Observatory,
 Mitaka, Tokyo 181-8588, Japan;kinoshita@nao.ac.jp}

\altaffiltext{4}{Department of Astronomy, Nanjing University,
Nanjing 210093, China;xhliao@nju.edu.cn}

\begin{abstract}
We have numerically explored the stable planetary geometry for the
multiple systems involved in a 2:1 mean motion resonance, and
herein we mainly concentrate on the study of the HD 82943 system
by employing two sets of the orbital parameters (Mayor et al.
2004). In the simulations, we find that all stable orbits are
related to the 2:1 commensurability that can help to remain the
semi-major axes for two companions almost unaltered over the
secular evolution for $10^{7}$ yr, and the apsidal phase-locking
between two orbits can further enhance the stability for this
system, because the eccentricities are simultaneously preserved to
restrain the planets from frequent close encounters. For HD 82943,
there exist three possible stable configurations:(1) Type I, only
$\theta_{1} \approx 0^{\circ}$, (2) Type II,
$\theta_{1}\approx\theta_{2}\approx\theta_{3}\approx 0^{\circ}$
(aligned case), and (3) Type III, $\theta_{1}\approx 180^{\circ}$,
$\theta_{2}\approx0^{\circ}$, $\theta_{3}\approx180^{\circ}$
(antialigned case), here the lowest eccentricity-type mean motion
resonant arguments are $\theta_{1} = \lambda _{1} - 2\lambda _{2}
+ \varpi_{1}$ and $\theta_{2} = \lambda _{1} - 2\lambda _{2} +
\varpi_{2}$, the relative apsidal longitudes  $\theta_{3} =
\varpi_{1}-\varpi_{2}=\Delta\varpi$ (where $\lambda _{1,2} $ are,
respectively, the mean longitudes of the inner and outer
planets;$\varpi_{1,2}$ are the longitudes of periapse). And we
find that the other 2:1 resonant systems (e.g., GJ 876 or HD
160691) may possess one of three stable orbits in their realistic
motions. In addition, we also propose a semi-analytical model to
study $e_{i}-\Delta\varpi$ Hamiltonian contours, which are fairly
consistent with direct numerical integrations. With the updated
fit, we then examine the dependence of the stability of this
system on the relative inclination, the planetary mass ratios, the
eccentricities and other orbital parameters: in the non-coplanar
cases, we find that stability requires the relative inclination
being $\sim 25^{\circ}$ or less; as to the planetary mass ratio,
the stable orbits for HD 82943 requires $\sin i \geq 0.50$ for a
fixed  value or $m_{1}/m_{2}\leq 2$ where $m_{2}$ remains for the
varying mass ratio; concerning the eccentricities, the system can
be always steady when $0 < e_{2}\le 0.24$ and $0 < e_{1}< 0.60$.
Moreover, we numerically show that the assumed terrestrial bodies
cannot survive near the habitable zones for HD 82943 due to the
strong perturbations induced by two resonant companions, but these
low-mass planets can be dynamically habitable in the GJ 876 system
at $\sim 1$ AU in the numerical surveys. Finally, we present a
brief discussion on the origin of the 2:1 resonance for HD 82943.
\end{abstract}


\section{Introduction}
The discovery of the extrasolar planets is opening a new world
beyond our solar system. Ever since 1995, Mayor \& Queloz (1995)
detected the first extrasolar giant Jupiter--51 Peg, and to date
there are more than 100 planetary systems discovered \footnote{see
also http://cfa-www.harvard.edu/planets/bibli.html and
http://exoplanets.org/}(Marcy, Cochran, \& Mayor 2000; Butler et
al. 2003) using the radial velocity technique in the surveys of
nearby young stars. At the time of writing, a dozen of multiple
planet systems--HD 82943, GJ 876, HD 168443, HD 74156, 47 Uma, HD
37124, HD 38529, HD 12661, 55 Cnc, $\upsilon$ And (Fischer et al.
2003), HD 169830 (Mayor et al. 2004) and HD 160691 (Jones et al.
2002) were discovered in recent years and this number is
undoubtedly cumulative as the measurements are being carried on.
Hence, it is necessary to categorize the discovered multiple
planetary systems according to their statistical characteristics
(such as the distribution of the planetary masses, semi-major
axes, eccentricities and metallicity) (Marcy et al. 2003), then to
study the correlation between mass ratio and period ratio (Mazeh
\& Zucker 2003) and further to improve the understanding of the
relationship between the planet occurrence rate and stellar
metallicity (Santos et al. 2003; Fischer, Valenti, \& Marcy 2004).
The other key point is to investigate possible stable
configurations for the multiple systems, in which the observations
reveal that most of them are typically characterized by mean
motion resonance (MMR) and (or) apsidal phase-locking between
their orbiting companions (Fischer et al. 2003; Ji et al. 2003a;
Lee \& Peale 2003), so that one can better understand the full
dynamics of these systems. In the present study, we mainly focus
our attention on the HD 82943 system.

As of April 4, 2001, the Geneva Extrasolar Planet Search Team
\footnote {see
http://obswww.unige.ch/$\sim$udry/planet/hd82943syst.html}
announced the discovery of the HD 82943 planetary system at an ESO
press release, consisting of two Jupiter-like planets orbiting the
parent star. Subsequently, Israelian et al. (2001, 2003) reported
the discovery of $^6$Li in the atmosphere of this metal-rich
solar-type star and indicated that the presence of $^6$Li can
probably be shown as evidence for a planet or planets having been
engulfed by the star HD82943. At first, let us review several
stellar features for this star:HD 82943 is a G0 star with
$B-V=0.623$, Hipparcos parallax of 36.42 mas, distance of 27.46
pc, and [Fe/H] = $0.29 \pm 0.02$ (Santos et al. 2001, 2003). The
star is aged 2.9 Gyr with the stellar mass of 1.15 $M_{\odot}$.
Similar to the GJ 876 system (Marcy et al. 2001), which is
believed that stability of the planetary system might be sustained
by the 2:1 mean-motion resonance (MMR) and apsidal alignment in
the semi-major axes (Kinoshita \& Nakai 2001; Lee \& Peale 2002;
Ji, Li, \& Liu 2002), the two planets of HD 82943 are now also
close to a 2:1 commensurability, with orbital periods $435.1 \pm
1.4$ and $219.4\pm 0.2$ \,d, and semi-major axes 1.18 and 0.75\,AU
(Mayor et al. 2004). However, the previous investigations
(Gozdziewski \& Maciejewski 2001) of the stability of the HD 82943
system showed that the system given by the earlier best-fit
orbital solution was extremely unstable  due to the strong
interaction of two massive planets with high eccentricities, hence
a natural question is that whether the two companions for this
system are really in a 2:1 orbital resonance and whether the
system is secularly stable with the observed best-fit orbital
parameters. Additionally, the stability for HD 82943 places
constraints on both planetary masses--if the mass of the outer
companion is smaller than that of the inner planet (Hadjidemetriou
2002), a system that harbors two planets moving on elliptic 2:1
resonant periodic orbits will be destabilized as a finale.
Nevertheless, the recently updated orbital solutions (Mayor et al.
2004), with more abundant observations as of October 13, 2003,
show that the inner planet is a bit more massive than the outer
one, thus it is indeed noteworthy to thoroughly reanalyze the
observed HD 82943 system, still these mentioned issues of this
fascinating system are to arouse our great interests and inspire
us to seek after the answers. Consequently, our first goal is to
carefully explore the system to fully understand in what kind of
likely configurations that two companions may orbit about their
host star and if exist, to reveal the dynamical mechanisms to
remain the system.

In a viewpoint of celestial mechanics, the former studies are
concerned for the dynamical analysis for the global stability of
the exosystems (Gozdziewski \& Maciejewski 2001, 2003; Ji et al.
2002; Kiseleva-Eggleton et al. 2002 ; Barnes \& Quinn 2004), the
investigations of resonant picture in the phase space
(Hadjidemetriou 2002; Ji et al. 2003a; Haghighipour et al. 2003;
Callegari Jr., Michtchenko, \& Ferraz-Mello 2004) or chaotic
behavior (Jiang \& Yeh 2004) in these systems, and the existence
of Habitable Zones (HZ) for Earth-like planets that supply stable
liquid-water environment to cultivate extraterrestrial intelligent
beings (Kasting, Whitmore, \& Reynolds 1993; Jones \& Sleep 2002;
Laughlin, Chambers, \& Fischer 2002; Dvorak et al. 2003; Menou \&
Tabachnik 2003). Other recent works on the 2:1 resonant geometry
of the extrasolar planetary systems include those by
Hadjidemetriou \& Psychoyos (2003), Beauge, Ferraz-Mello, \&
Michtchenko (2003) and Lee (2004). They studied the presence and
location of stable equilibrium solutions or dynamical evolution of
such exosystems in more general way. In this work, we revisit the
dynamics of the HD 82943 system and outline several possible
stable configurations for a system harboring two companions in a
2:1 resonance, and our new findings are that this system can be
locked in aligned or antialigned orbital motions, which these
results are never presented nor analyzed in other studies.
Specifically, at first, we introduce the numerical setup (see \S2)
for the dynamical simulations. In \S3, we make a study of the
various planetary configurations for HD 82943 by means of the
semi-analytical treatment and direct integrations over long-term
orbital evolution. We present the numerical results of HD 82943 by
employing two series of the best-fit orbital solutions, and we
discover three kinds of stable planetary geometry for this system
that all are linked to a 2:1 resonance. Furthermore, we advance a
semi-analytical model that can avoid the difficulties in  the
perturbation expansions for the larger eccentricities and still
help to explain the numerical evolutions. Still, we investigate
how the stability for the resonant topology depends on the
planetary mass ratio and the orbital parameters on the basis of
the new fit (Mayor et al. 2004). In \S4, we explore whether there
exist an Earth-like planet surviving about the Habitable zones for
the authentic systems of HD 82943 and GJ 876.

The other extraordinary phenomenon in the exosystems is that many
of them can host giant Jupiter-like planets with strikingly larger
eccentricities, and in Figure 1 is shown the distribution of the
eccentricities of 111 planets\footnote{The data were taken from
http://exoplanets.org, as of Aug. 5, 2003} discovered at present
day. The figure exhibits that more than 50\% of the planets have
the eccentricities larger than 0.30, and HD 80606 b (Naef et al.
2001) can occupy the eccentricity amounting up to 0.93, which
indicates that such unusual discoveries are quite different from
the circumstances in our solar system, where most of the major
planets revolve around Sun on the near-circular orbits. In
addition, from the view of the observations, the larger
eccentricities seem to favor the planet detectability, where the
semi-amplitude of wobble velocity $K \propto {M_{p} \sin i}/
{\sqrt{a(1-e^2)}}$ (with $M_{p}\ll M_{c}$), herein $M_{c}$,
$M_{p}$, $a$, $e$ and $i$ are, respectively, the stellar mass, the
planetary mass, the orbital semi-major axis, the eccentricity and
the inclination of the orbit relative to the sky plane, requires
more massive planetary mass ($\sim$ the order of Jupiter's mass),
a lower $a$ (e.g., close-in giant planets) and a higher $e$ (see
Fig. 1). Moreover, many previous authors have put forward diverse
theories and possible mechanisms to interpret the evolution and
origin of the orbital eccentricity variation in the exosystems:
the disk-planet or planet-planet interaction (Goldreich \&
Tremaine 1980; Weidenschilling \& Marzari 1996; Lin \& Ida 1997;
Ford, Rasio, \& Sills 1999) render the orbital migration of gas
giant planets embedded in the protoplanetary systems (Lin,
Bodenheimer, \& Richardson 1996; Ward 1997; Bryden et al. 2000;
Nelson et al. 2000) and the eccentricities are usually believed to
excite to the observed values through complicated processes
(Snellgrove, Papaloizou, \& Nelson 2001; Kley 2003) or hybrid
mechanisms (Lee \& Peale 2002; Chiang, Fischer, \& Thommes 2002)
related to both the planetary migration and the resonant capture.
Therefore, as a second objective, we intend to preliminarily make
a discussion of the possible origin of the eccentricities or
resonant configurations (see \S5) for two planets of HD 82943. As
a final part, in \S6, we summarize our principal results and give
a brief discussion.

\section{Numerical setup}
In the present work, we aim to numerically investigate the orbital
motions for the two companions for HD 82943 in a three-dimensional
space. And at first, let us bear in mind that the two companions
of this system are assumed to be in the same orbital plane for our
simulations except where noted. Here we adopt the best-fit orbital
parameters from  the web site of the Geneva Team.  As is known,
the two-Kepler fit produce five parameters set-($K_{i}$, $P_{i}$,
$e_{i}$, $\omega_{i}$, $T_{pi}$), where for each planet $i=1,2$,
are the amplitude $K_{i}$, the orbital period $P_{i}$, the
eccentricity $e_{i}$, the argument of periapse $\omega_{i}$ and
the time of periapse passage $T_{pi}$. The Geneva Team presented
two sets of the orbital data and hereafter we respectively call
them Fit 1 and Fit 2. Here for Fit 1 of 107 observations (as of
July 31, 2002) with the residual of 7.4 m/s, the data are listed
here: $m_{1} \sin i = 0.88$ $M_{Jup}$, $m_{2} \sin i = 1.63$
$M_{Jup}$, $a_{1} = 0.73$ AU, $a_{2} = 1.16$ AU,
$e_{1}=0.54\pm0.05$, $e_{2}=0.41\pm0.08$, $\omega_{1}=138\pm
13^\circ$ and $\omega_{2}=96\pm 7^\circ$, where hereafter the
subscripts 1 and 2 denote the inner and outer planets,
respectively. And Fit 2 were derived by fitting 142 observations
(as of October 13, 2003, see also Mayor et al. 2004) with the
residual of 6.8 m/s, the orbital parameters are presented: $m_{1}
\sin i = 1.85$ $M_{Jup}$, $m_{2} \sin i = 1.84$ $M_{Jup}$, $a_{1}
= 0.75$ AU, $a_{2} = 1.18$ AU, $e_{1}=0.38\pm0.01$, $e_{2}=0.18
\pm0.04$, $\omega_{1}=124\pm 3^\circ$ and $\omega_{2}=237\pm
13^\circ$. Apparently, the eccentricities, semi-major axes,
periapse arguments, true anomalies and  others can act as key
factors to determine the shape of the radial velocity curves.
However, due to the large interaction between two planets in a low
mean motion resonance, Laughlin \& Chambers (2001) pointed out
that short-term perturbations among massive planets in multiple
planet systems (such as GJ 876, HD 82943) can result in radial
velocity variations of the central star that differ substantially
from velocity variations derived assuming the planets are
executing independent Keplerian motions. On the other hand, the
best-fit orbital solutions can vary a little from time to time
provided that more updated observations were supplied or other
elaborate fitting procedure was considered (Mayor et al. 2004), or
the velocity trend is removed due to the unveiling of an
additional planet or more planets (Fischer et al. 2001) for a
specific system. Therefore, at this stage, it is meaningful to
study the dynamical characteristics of the system based on the
subsequent points: (i) slightly varying the orbital parameters
about the best-fit parameters by adding the uncertainties with
direct integrations, (ii) seeking for the possible stable geometry
for HD 82943 in parameter space, (iii) the dynamical mechanisms of
retaining the system and (iv) the likely eccentricity or
configuration origin for two planets, which motivate us to
dedicate to this contribution.

Next, we continue to describe the means of preparation for the
initial orbital data before we start the numerical integrations:
for each planet, we usually need to obtain six orbital elements -
the semi-major axis $a$, the eccentricity $e$, the orbital
inclination $I$, the nodal longitude $\Omega $, the argument of
periastron $\omega $ and the mean anomaly $M$. Here, two groups of
the initial orbital elements were produced on the basis of the
data by Fit 1 and Fit 2. In  each group, we assume that the
semi-major axes of the two planets are always unchanged, say, for
Fit 2, being 0.75 and 1.18\,AU respectively. As we have indicated
above, the two resonant planets are supposed to be coplanar for
the initial configuration and the inclinations are both taken as
small constant values not far from zero, e.g., $0.5^{\circ}$. The
eccentricities and arguments of periapse are generated in the
orbital parameters space in the proximity of the best-fit orbital
solutions given the nominal observation errors. For example, for
Fit 2, we take the observed eccentricities $e_{1}$ or $e_{2}$ to
be centered, respectively, at 0.38 and 0.18, and randomly
displaced by not more than $3\sigma$, respectively, then we obtain
the resulting initial eccentricities for integration. Still, we
carried out similar steps to achieve the starting values of
periapse arguments. The remained two angles of nodal longitudes
and mean anomalies\footnote{Laughlin G. and Lee M.H. (2003,
private communications) pointed out to us that the mean anomalies
and periapse arguments should perfectly match the radial
velocities. Here, we choose proper values for $M$, $\omega$ and
$\Omega$, such that the linear combination of initial arguments
can be close to the resonant configuration, this is reasonable,
because the real system is observably stable and approximate to
2:1 MMR. And in Table 1 and 2, we present the orbital solutions
where the measurement errors of $\Delta\omega=0$ and $\Delta e=0$,
indicating that the eccentricities and periapse arguments are just
equal to those by the Geneva Team.} are randomly distributed
between $0^{\circ}$ and $360^{\circ}$. As a consequence, we
obtained 100 orbits for each planet, to perform the integration of
the planetary system, and each pair of the orbits was integrated
for the time span of $10^7$ yr, unless otherwise stated.
Obviously, these orbits just lie in the neighborhood of the
fitting parameters and represent the orbital motions quite close
to the reality. Here we make study of the possible motion near the
best-fit orbital solutions and expect that such investigations can
be helpful to reveal some important dynamical features for the
studied system or present valuable clues on the searching for
other exosystems related to a 2:1 resonance.

With the N-body codes (Ji et al. 2002), we carried out the
numerical integrations for the HD 82943 system. In our
simulations, we adopted the mass of the host star to be $1.05$ $
M_{\odot}$ for Fit 1, and $1.15$ $ M_{\odot}$  for Fit 2. Under
the assumption of sin\textit{i} = 1, the minimum masses of two
companions are employed throughout the paper unless we state
elsewhere. In addition, we utilized the time step to be one
percent of the orbital period of the inner planet (HD 82943 c) in
the integrations. The numerical errors were effectively controlled
during the integration, with the local truncation error $10^{-14}$
for the time span of 10$^{7}$ yr, in the meantime the accuracy was
also checked by the energy errors. In comparison, we also employed
symplectic integrators (Feng 1986; Wisdom \& Holman 1991) to
integrate the same orbit to assure the results for some cases.

In the research of the planetary systems, the maximum Lyapunov
characteristic exponent (LCE) is usually adopted to identity the
regular or chaotic orbits, since the chaotic orbits are
sensitively dependent on the initial conditions, corresponding to
a positive LCE in the secular evolution. In this paper, we refer
to the stability for the system means that the orbiting planets
can remain periodic or quasi-periodic motions with bounded
trajectories after the investigated time was done. We define the
unstable orbits that either of the planets is ejected far away or
moves too close to their parent star, and in our simulations, the
integration was automatically ended when meeting the following
criteria: (1) either of the eccentricities approaches unity, (2)
either of the semi-major axes exceeds the factor of 2 or reduces
half of the starting values, (3) either of the planets collides
with the star or these planets do each other when entering the
scope of the mutual Hill sphere.

\section{Stable planetary geometry in 2:1 mean motion
resonance} To begin the study of the stable resonant geometry in
HD 82943, let us firstly recall several known facts in our solar
system. One may be aware that the pair of Jupiter and Saturn is in
a near 5:2 commensurability. In general, the mean motion resonance
takes place in the pairs of the moons of the major planets (such
as the Galilean satellites of Jupiter: Io-Europa-Ganymede; see Lee
\& Peale 2002), the asteroidal belt and  Kuiper Belt Objects
(KBOs) (Duncan, Levison, \& Budd 1995; Wan \& Huang 2001). As more
and more extrasolar planetary systems are being discovered, the
resonant pairs are frequently found to occur in the multiple
systems, e.g., HD 82943 (Gozdziewski \& Maciejewski 2001), GJ 876
(Lee \& Peale 2002), and possibly HD 160691 (Gozdziewski, Konacki,
\& Maciejewski 2003; Bois et al. 2003) in a 2:1 MMR, and 55 Cnc in
a 3:1 MMR (Ji et al. 2003a), and so on. Several theoretical or
numerical works have been done to enhance the comprehension of the
evolution into the resonances by non-conservative tidal forces
(Ferraz-Mello, Beauge, \& Michtchenko 2003), the convergent
migration for the planets due to the disk-planet interaction by a
damping force (Snellgrove et al. 2001; Lee \& Peale 2002; Kley
2003), the excitation of the orbital eccentricities by repeated
crossings of two companions migrating on divergent orbits (Chiang
et al. 2002; Chiang 2003) or the eccentricity growth by two
resonant planets through inward migration (Murray, Paskowitz, \&
Holman 2002), and resonant inclination excitation mechanism for
migrating planets in the gas disk (Thommes \& Lissauer 2003). On
the other hand, if the two planets are trapped into a 2:1
resonance via one of the potential mechanisms, it is still
necessary to understand their actual situation of orbital motions
after the capture, and to investigate whether they can secularly
survive in the system in some appropriate  geometry and examine
how the stability depends on the orbital parameters.

In the usual notation of celestial mechanics, the lowest order
eccentricity-type (see Murray \& Dermott 1999) resonant arguments
$\theta_{1}$, $\theta_{2}$ for the 2:1 MMR are
\begin{equation}
\label{eq1} \quad\quad\theta_{1} = \lambda _{1} - 2\lambda _{2} +
\varpi_{1},
\end{equation}
\begin{equation}
\label{eq2} \quad\quad\theta_{2} = \lambda _{1} - 2\lambda _{2} +
\varpi_{2},
\end{equation}
\noindent where $\lambda _{1} $, $\lambda _{2} $ are ,
respectively, the mean longitudes of the inner and outer planets,
and $\varpi_{1}$, $\varpi_{2}$ denote their apsidal longitudes,
respectively. Additionally, the relative apsidal longitudes of two
companions $\theta_{3}$ reads,
\begin{equation}
\label{eq3}
\theta_{3} = \varpi_{1}-\varpi_{2}=\Delta\varpi .
\end{equation}
As we mentioned previously, the  apsidal alignment (or
antialignment) is found to exist in  most of the multiple systems
(Kinoshita \& Nakai 2000; Rivera \& Lissauer 2000; Chiang,
Tabachnik, \& Tremaine 2001; Laughlin et al. 2002; Lee \& Peale
2002, 2003; Gozdziewski et al. 2003; Ji et al. 2003a, b; Zhou \&
Sun 2003), to play a  significant role in retaining the stability
of the system (Gozdziewski 2003; Ji et al. 2003b). Moreover, this
apsidal phase-locking of two orbits indicates that a pair of
planets have common time-averaged rate of apsidal precession, and
in the literature, it is also referred to apsidal resonance or
corotation (Chiang et al. 2001; Malhotra 2002). For a 2:1 MMR, it
is easily noticed that for the above three arguments, no more than
two are linearly independent, indicating that either all three
librate or only one librate in the occurrence of the libration of
the critical argument. In more general case, Nelson \& Papaloizou
(2002) also showed that both situations can occur for a $\it{p :
q}$ eccentricity-type commensurability (Murray \& Dermott 1999),
where the associated resonant angles are defined by $\phi_{p,q,k}
(\lambda_{1},\lambda_{2},\varpi_{1},\varpi_{2}) = p\lambda_{1} -
q\lambda_{2} + (k-p)\varpi_{1} + (q-k)\varpi_{2})$($p, q, k$ are
positive integers and $p \le k \le q$) in the numerical surveys of
the pair of the extrasolar planets. As a matter of fact, this was
again verified in the case of the 3:1 MMR for 55 Cnc (Ji et al.
2003a; Zhou et al. 2004), where not only do all three resonant
angles of $\lambda_{b}-3\lambda_{c}+2\varpi_{c}$,
$\lambda_{b}-3\lambda_{c}+\varpi_{b}+\varpi_{c}$ and
$\lambda_{b}-3\lambda_{c}+2\varpi_{b}$ (here the subscript b, c,
respectively, denote the two inner planets of 55 Cnc b and 55 Cnc
c) librate, but the relative apsidal longitude of $\Delta\varpi$
librates about a constant angle simultaneously.

In the simulations of HD 82943, we found that three types of
stable orbits can survive in this system: (I) Only $\theta_{1}$
librates about $0^{\circ}$, while $\theta_{2}$ and $\theta_{3}$
circulate; (II) Case of alignment, three arguments all librate
about $0^{\circ}$, denoting by
$\theta_{1}\approx\theta_{2}\approx\theta_{3}\approx 0^{\circ}$;
(III) Case of antialignment, both $\theta_{1}$ and $\theta_{3}$
librate about $180^{\circ}$, while $\theta_{2}$ librates about
$0^{\circ}$, denoting by $\theta_{1}\approx 180^{\circ}$,
$\theta_{2}\approx0^{\circ}$, $\theta_{3}\approx180^{\circ}$. In
the following sections, we will separately discuss above three
stable configurations with both the numerical outcomes and
analytical means and also compare them with remarkable evidences
in other multiple systems involved in 2:1 resonance.

\subsection{Coplanar semi-analytical model}
First, let us suppose two planets with the masses of $m_{1}$ and
$m_{2}$ orbit the central star with the mass of $m_{0}$. Here we
only take into account the point-mass interaction between
star-planet and planet-planet, without the consideration of the
effects of the oblateness and  general relativity  arising from
the parent star. Thus, in Jacobi coordinates ($a_{i}, i=1, 2$;
assume $a_{1}<a_{2}$), the Hamiltonian for three-body system can
be written (Brouwer  \& Clemence 1961):
\begin{equation}
  F = F_{0} + \Phi,
\end{equation}
where
\begin{equation}
  F_{0} = - \frac{Gm_{0}m_{1}}{2a_{1}} -
  \frac{Gm_{0}m_{2}}{2a_{2}},
\end{equation}
and
\begin{equation}
 \Phi = - \frac{G m_{1}m_{2}}{r_{12}} -
 Gm_{0}m_{2}\left(\frac{1}{r_{02}}-\frac{1}{r_{2}}\right).
\end{equation}
Here $r_{02}$ (or $r_{12}$) is the distance between $m_{0}$ (or
$m_{1}$) and $m_{2}$, $r_{2}$ the distance of $m_{2}$ with respect
to the center of mass for $m_{0}$ and $m_{1}$. Therefore, the
disturbing potential $\Phi$ can be further expanded to the
following forms for the coplanar case,
\begin{equation}
\label{eq7}
 \Phi = - \frac{G m_{1}m_{2}}{a_{2}}\sum
 S(a_{1},a_{2},e_{1},e_{2})\cos \phi ,
\end{equation}
where
\begin{equation}
 \phi = k_{1}\lambda_{1} + k_{2}\lambda_{2}+ k_{3}\varpi_{1} +
 k_{4}\varpi_{2},
\end{equation}
Here
\begin{equation}
 \sum_{i=1}^{4}k_{i}=0.
\end{equation}
In  (\ref{eq7}), $S$ depends on the semi-major axes (in
association with the Laplace coefficients) and the eccentricities
of two planets and this expansion of the disturbing potential also
requires the smaller values of $e_{i}$ (i=1,2) and the two orbits
without intersection. In the case of GJ 876, Lee \& Peale (2002)
expanded (\ref{eq7}) and collected the terms up to $e^3$, the
truncation series may be valid for low values of $e_{i}$, where
$e_{1}=0.255$ and $e_{2}=0.035$ (Laughlin \& Chambers 2001),
without the occurrence of the crossing orbits. In a more recent
work, Beauge \& Michtchenko (2003) also developed an expansion
method that can deal with higher values of the eccentricities up
to a limit of $e_{i}\sim 0.5$, but they suggested that it is still
related to the situation of the proximity between the planets.
However, as for HD 82943, the planetary eccentricities can be as
high as 0.41-0.54 for Fit 1, or 0.18-0.38 for Fit 2, the
development of (\ref{eq7}) with respect to large eccentricities
cannot be exactly applied, because the expanding series,
consisting of a great number of the expansion items, either
converge very slowly or become divergent in the collision
circumstances. Therefore, herein we seek for a semi-analytical
treatment (Kinoshita \& Nakai 2002) for this question and then
compare the analytical results with direct numerical integrations.
We emphasize that this semi-numerical method  corresponds to a
first-order theory with respect to the small parameter $\epsilon$,
where $\epsilon$ represents  the mass ratio of the Jupiter-like
planet to its parent star and is normally of the order of $\sim
10^{-3}$ in the planetary systems, and can apply to the planetary
system with any eccentricities, even for the mutual crossing case.

The Hamiltonian for the coplanar case is
\begin{equation}
\label{eq8}
F=F(a_1, a_2,e_1,e_2,\varpi_1,\varpi_2,\lambda_1,\lambda_2).
\end{equation}
In order to keep the Hamiltonian form, we should use Jacobi
coordinates to study this system. As the indirect part of the
Hamiltonian $F$ does not contribute to the secular part, we simply
take the direct part. As the two orbits of HD 82943 are close to
each other and may intersect (Ji  et al. 2003c; Hadjidemetriou \&
Psychoyos 2003; Lee  2004), the ordinary analytical expansions of
the main part invalidate in the practical usage, then we adopt the
original form of the main part of the disturbing function and
numerically evaluate it. Again, by eliminating short-periodic
terms in the classical average process, the new Hamiltonian reads:
\begin{equation}
\label{eq9} F^*=F^*(a_1,a_2,e_1,e_2,\theta_1,\theta_3).
\end{equation}
The degree of freedom of the new Hamiltonian (\ref{eq9}) is
reduced from four to two. However, the Hamiltonian $F^*$ is not
integrable. As we shall see in \S3.2 that the semi-major axes of
the two resonant planets slightly change when they are close to
2:1 MMR, so we can assume that $a_1$ and $a_2$ are almost constant
(see Figures 2 and 3) for the critical argument
$\theta_1=0^{\circ}$ or $\theta_1=180^{\circ}$. In final, the
degree of freedom of the new Hamiltonian is reduced to one:
\begin{equation}
\label{eq10} F^*=F^*(e_1,e_2,\theta_3).
\end{equation}
\noindent We again eliminate $e_2$ in equation (\ref{eq10}) with
conservation of the total angular momentum $J$ (see also
Appendix), then we have
\begin{equation}
\label{eq13} F^*=F^*(e_1, \theta_3, J).
\end{equation}
\noindent Thus we can draw the level curves of the Hamiltonian and
understand the global behavior of $e_1$ and $\theta_3$. As the
eccentricity $e_1$ of the inner planet becomes large, we
numerically averaged the original Hamiltonian (\ref{eq8}) under
the condition of the critical argument $\theta_1=0^{\circ}$ or
$\theta_1=180^{\circ}$ and the angular momentum conservation
relationship, and then obtain numerically the averaged Hamiltonian
(\ref{eq13}) with the parameter of the total angular momentum $J$.
The technique of the numerical averaging is extensively described
in the paper (Kinoshita \& Nakai 1985), in which the case of
$\theta_1\neq 0$ is also discussed. To make a study of the
evolution of the eccentricities and the relative periapse
arguments, we can draw the contour map of Hamiltonian (\ref{eq13})
by taking $\theta_3$ as the horizontal axis and $e_1$ or $e_2$ as
the vertical axis (see Figure 4) with the parameter $J$, which is
determined from the initial conditions.

\subsection{Results for Fit 1}
For the first 100 runs for Fit 1, each orbit integration lasts for
$t=10^7$ yr. In the simulations, we notice that all the stable
cases are involved in the 2:1 MMR and easily understand that the
stability of a system is sensitive to its initial planetary
configuration. Furthermore, we noticed that 2\%, 3\%, and 2\% of
the cases belong to Types I, II, and III and 93\% of the systems
destabilize, which means that either of the two planets leave
\textit {in  situ} to the infinity with a rapid increase of the
semi-major axes or the eccentricities grow much larger on the
timescale $\sim10^{4} - 10^{5}$ yr or even shorter, owing to the
mutual interaction  between star-planet or planet-planet through
frequent close approaches. In Table 1, are listed two sets of
orbital parameters--Fit 1a is related to the antialigned
configuration for Type III and Fit 1b is involved in the case for
aligned orbits for Type II. Here the adopted parameters for two
planets are the masses of 1.63-0.88 $M_{jup}$, the semi-major axes
of 1.16-0.73 AU, the eccentricities of 0.41-0.54. Let us first
examine the stable orbits of Type I, where in the numerical
simulations, we find that only one argument of $\theta_{1}$
librates about $0^{\circ}$ and the others $\theta_{2}$ and
$\theta_{3}$ take up the full circulation from $0^{\circ}$ to
$360^{\circ}$. The semi-major axes $a_{1}$ and $a_{2}$ perform
tiny oscillations about 0.73 and 1.15 AU  for the whole
integration time. Type I was initially introduced by Gozdziewski
\& Maciejewski (2001) with the previous fits. However, in this
paper, Types II and III, which can grasp the couple of companions
escaping from deviating their tracks in two directions of
remaining both the semi-major axes and the eccentricities, are new
findings of stable orbits for this resonant system. Hence, in the
following sections, we primarily pay attention to the stable
configurations linked to both a 2:1 resonance and the apsidal
corotation.

\subsubsection{Antialigned  case}
Figure 2 shows the  typical orbital evolution for Fit 1a
(antialigned orbits). In the figure, we can see that the
semi-major axes $a_{1}$ and $a_{2}$ slightly vibrate about 0.73
and 1.15 AU for $t=10^{7}$ yr \footnote{We also extend the
integration to $10^8$ yr using symplectic integrator for this
case, and no sign shows that the system will become chaotic for
longer evolution.} (in the figure, to see more clearly, we simply
display a snapshot for $t=2000$ yr), meanwhile $\theta_{1}$
librates about $180^{\circ}$ with a moderate amplitude of $\sim
30^{\circ}$, $\theta_{2}$ about $0^{\circ}$ with a smaller
amplitude of $\sim 10^{\circ}$ and $\theta_{3}$ (by \textit{thick
line}) about $180^{\circ}$ with an amplitude of $\sim 30^{\circ}$.
It is noteworthy to point out that the 2:1 resonance variable
$\theta_{1}$ librates in accord with the relative apsidal
longitudes $\theta_{3}$, occupying the same period of $\sim$ 300
yr. This phenomenon can be easily understood because the above
three angles are not independent in the course of the orbital
motion. Thus we can call $\theta_{1}$ and $\theta_{3}$ as two
fundamental variables to characterize the stable planetary
geometry in 2:1 MMR. In addition, the modulations of the
semi-major axes $a_{i}$ and  $\theta_{i}$ ($i=1,2$) reveal the
fact that the two planets are indeed trapped in a 2:1 MMR for the
secular orbital evolution. In Figure 2, we further notice that the
eccentricities of $e_{1}$ and $e_{2}$ separately range in (0.36,
0.56) and (0.39, 0.49), bearing the equal libration periods of
$\theta_{1}$ and $\theta_{3}$. The orbital antialignment of axes
in HD 82943 is reminiscent of that in the 2: 1 orbital resonances
for Io-Europa system (Lee \& Peale 2002), in which $\theta_{1}$
(involved in the Io's longitude of periapse) librates about
$0^{\circ}$, and $\theta_{2}$ (involved in the Europa's longitude
of periapse , $\theta_{3}$ librate about $180^{\circ}$,
respectively. The antialigned circumstance for Io-Europa is
equivalent to Type III in the sense of dynamically stable
configuration, however, these jovian moons are occupying almost
near-circular orbits resulting from the primordial or the tidal
origin (Peale 2003) and the resonance configuration may come into
reality when the eccentricities are small. By contrast, the HD
82943 system differs the Io-Europa pair in that both of the
planets possess high eccentricities at present day. Hence, the
origin of the 2:1 resonance for HD 82943 (Fit 1a) orbiting two
massive planets with elliptic trajectories should be explained by
an innovative mechanism rather than a capture induced by
co-orbital differential migration (S. Peale 2003, private
communication; see also \S5) and perhaps stem from gravitational
scattering in a crowded system that augment their eccentricities,
i.e., for a system with 10 planets (Adams \& Laughlin 2003), the
scattering processes can yield the full extent of possible
eccentricity where $0 \lesssim e \lesssim 1$.

Recently, Bois et al. (2003) found that the stable configuration
of the HD 160691 system is possibly involved in a 2:1 mean motion
resonance combined with an anti-aligned in two apsidal lines, with
higher eccentricity ($e_{2} > 0.52$) of the outer planet.
Obviously, their research supported the stable geometry of Type
III of HD 82943, which also implies that this resonant topology
may be suitable especially for the planets bearing larger
eccentricities.

Moreover, this antialigned resonant configuration for Fit 1a means
that the conjunctions can take place when the inner planet  moves
near apoapse and the outer planet is close to the periapse, and
\textit{vice versa}, which indicates that the two planets are far
from each other during their orbital evolution and protected from
frequent close encounters. Therefore, we can safely conclude that
the stability of the HD 82943 system, which is related to the
above coplanar configuration for Fit 1a, can be simultaneously
sustained by two dynamical mechanisms: the 2:1 MMR and the apsidal
antialignment.

\subsubsection{Aligned  case}
Figure 3 exhibits the orbital evolution for Fit 1b (aligned
orbits). Consequently, let us note that the semi-major axes
$a_{1}$ and $a_{2}$ do not change dramatically but undergo small
oscillations about 0.73 AU and 1.15 AU for $t=10^{7}$ yr (a
snapshot for $t=3000$ yr), further the eccentricities $e_{1}$ and
$e_{2}$ librate in the extent of (0.5, 0.8) and (0, 0.45),
respectively, in the behavior of the converse cycles, owing to the
conservation of the total angular momenta for the system. Here
$\theta_{1}$ librates about $0^{\circ}$ with a small amplitude of
$\sim 20^{\circ}$, but $\theta_{2}$ and $\theta_{3}$ (by
\textit{thick line}) individually librate about $0^{\circ}$ with a
large amplitude of $\sim 70^{\circ}$. In this resonant aligned
geometry, we again observe that $\theta_{1}$, $\theta_{2}$ and
$\theta_{3}$ share the common librating period of $\sim 700$ yr,
in the same time, coupled with the librating period of the
eccentricities of $e_{1}$ and $e_{2}$. In addition , it is
worthwhile to point out that  at the time that $\theta_{1}$,
$\theta_{2}$ and $\theta_{3}$ all approach zero (where
$\theta_{1}$ and $\theta_{3}$ start from the negative values to
zero), the eccentricity of $e_{1}$ reaches the maximum $\sim 0.8 $
and $e_{2}$ is close to the minimum $\sim 0 $, thus at the aligned
conjunction point, the separation between two planets $d = a_{2}(1
- e_{2}) - a_{1}(1 - e_{1})$ (where $a_{1}$ and $a_{2}$ are almost
preserved  on the condition of 2:1 resonance locking) can reach
the maximal value and the maximum separation can prevent them from
strong mutual perturbation when they rendezvous at the synodic
location. Similar analysis can be applied for the case of Fit 1a,
where both $\theta_{1}$ and $\theta_{3}$ run through $180^{\circ}$
(see Figure 2). Again, let us make a brief comparison with the
circumstance for the 2:1 resonant system GJ 876, which is also
found to be in aligned configuration for the two orbits (Marcy et
al. 2001). Lee \& Peale (2002) further found that the amplitude of
the librations  of three arguments are not far from $0^{\circ}$ by
using Laughlin-Chambers solutions and they even indicated that all
the librations are remained even for the amplitudes of
$\theta_{1}$ amounting up to  $45^{\circ}$ for some cases. As for
Fit 1b, let us bear in mind the fact that the planetary
eccentricities are much larger than those for GJ 876, although
they may have similar planetary orbital geometry. Sequently, the
conjunctions always take place when two planets move close to
their the periapse longitudes and the alignment can be maintained
over long-term orbital evolution. This, \textit {de facto},
reveals that Type II coplanar configuration may be possible for
real observed systems of GJ 876 and HD 82943.

\subsubsection{Comparison with the semi-analytical results}
Here we adopted the semi-analytical model introduced in \S3.1 to
draw the Hamiltonian contour and then compare these results with
those given by direct integrations over secular orbital evolution.
In the averaged Hamiltonian after removing the short-periodic
terms, the difference arising from the choice between the Jacobi
coordinates and  astrocentric coordinates is the order of the
perturbation ($m_1/m_0, m_2/m_0$), which could be ignorable.
Therefore, we adopt the astrocentric elements given in Table 1 to
be Jacobi elements in the averaged model. As a comparison, the
initial orbital elements for numerical integration are also
converted into the Jacobi coordinates. In Figure 4, we indicate
that the contour diagram of Hamiltonian (\ref{eq13}) are plotted
against various levels by changing $e_{1}$, $e_{2}$ and
$\Delta\varpi=\varpi_{1}-\varpi_{2}$, \textit {left panel}
represents the results for Fit 1a, where the contour levels are
exhibited by \textit{thin line} and the numerical solutions shown
by \textit{thick line} (the innermost curve about marked
\textit{plus}), while \textit {right panel} for Fit 1b. The
$e_{i}-\Delta\varpi$ figures show a good agreement between the
numerical outcomes and the semi-analytical solutions for two fits.
It is worthwhile to mention that the closest level of the
Hamiltonian contour to the numerical solution set the stable
boundary for the apsidal resonant geometry, inside which the
regular orbital motion for two orbits can be found with qualified
eccentricities and the relative apsidal longitudes, thus we may
conclude that this semi-analytical means is effective to help
predict stable orbital solutions (e.g., Beauge et al. 2003). On
the other hand, the diagrams of the Hamiltonian contour further
show that the stability zones for the eccentricities $e_{1}$ and
$e_{2}$ are connected with the librating amplitude of
$\Delta\varpi$: the smaller stability zones of the eccentricities,
the narrower modulations of the relative apsidal longitudes, and
\textit{vice visa}.

In addition, we calculated the equilibrium points for each fit:
($180^{\circ}$, 0.45) and ($180^{\circ}$, 0.46), respectively, for
Fit 1a; ($0^{\circ}$, 0.68) and ($0^{\circ}$, 0.29), respectively,
for Fit 1b.  As for Fit 1b, we observe that there should exist
stable orbits about the equilibrium center for $(\Delta\varpi,
e_{2})=(0^{\circ}, 0.29$), together with high eccentricity $e_{1}$
above 0.60, which satisfy the necessary conditions for two planets
evolving into a 2:1 resonance (Kley, Peitz, \& Bryden 2004; see
also \S5). Again, we recall that the period of the 2:1 resonant
variable $\theta_1$ is equal to that of the relative apsidal
longitudes $\theta_3$ (see Figs. 2 and 3), which can be clearly
reflected in Fig. 4. Moreover, in a view of the numerical results,
we can see that the eccentricities of both planets are well
restricted with the small-amplitude libration of $\theta_3$ about
$180^{\circ}$ for Fit 1a (antialigned) and large-amplitude
libration of $\theta_3$ about $0^{\circ}$ for Fit 1b (aligned),
which can also be consistent with the results in Figs. 2 and 3.

As aforementioned, different selections for the initial orbital
parameters will lead to stable or unstable configurations for the
HD 82943 system. By  analyzing the starting orbital data, we
discover that those stable planetary orbits involved in Type II
(or III) require the initial $\theta_1$ and $\theta_3$ should be
constructed to satisfy the conditions of being less than tens of
degrees from $0^{\circ}$ (or $180^{\circ}$), implying that the
linear combinations of the mean anomalies, ascending nodes and
periapse arguments can be close to the 2:1 resonant geometry.

\subsection{Results for Fit 2}
As for the second 100 runs near the vicinity of Fit 2, we found
there exist two kinds of stable geometry related to a 2:1 MMR for
this new fit (Mayor et al. 2004), in which 16\% of the stable
systems can last for $10^{7}$ yr (where 14\% of the stable cases
is in aligned orbits of Type II and 2\% with simply $\theta_{1}$
libration for Type I), and 84\% are found to self-destruct at $ t
< 10^{6}$ yr. The percentage of the survival orbits indicates that
there is a likelihood for the two planets of this system to be
aligned. In Table 2, are listed the orbital parameters in
association with the alignment of two orbits. Let us recall the
adopted planetary masses of 1.84-1.85 $M_{jup}$, the semi-major
axes of 1.18-0.75 AU and the eccentricities of 0.18-0.38. In the
following subsections, the additional computations are carried out
most frequently according to the initial orbits from Fit 2, and
the typical secular evolution timescale ranges from $10^{6}$ yr to
$10^{7}$ yr.

Figure 5 illustrates the typical aligned orbital evolution  for
Fit 2. In  comparison with Figure 3, we notice that the semi-major
axes $a_{1}$ and  $a_{2}$ are almost unchanged but modulate about
0.75 and 1.18 AU  with small amplitude for $t=10^{7}$ yr, at this
time, the amplitude of the oscillations for $e_{1}$ and $e_{2}$
are not so large and the eccentricities are just wandering in the
span (0.34, 0.44) and (0, 0.25), respectively. Here $\theta_{1}$
librates about $0^{\circ}$ with a moderate amplitude of $\sim
45^{\circ}$ (in coincidence with one of the aligned case for GJ
876 found by Lee \& Peale 2002; see also \S3.2.2), but
$\theta_{2}$ and $\theta_{3}$ (by \textit{thick line})
individually librate about $0^{\circ}$ with large amplitude of
$\sim 80^{\circ}$. Again, we notice that the eccentricities have
the libration periods of $\sim$ 600 yr, which are coupled with
those of  $\theta_{1}$ and $\theta_{3}$. It should mention that
although the various starting orbital parameters between Fit 2 and
Fit 1b can result in variational librating amplitudes in the
eccentricities and resonant arguments, however, from a viewpoint
of topology, there is not so much difference between two aligned
orbital solutions. Furthermore, in Figure 6, we still plot
$e_{i}-\Delta\varpi$ Hamiltonian contour (by \textit{thin lines})
and the numerical results are presented by dotted points. The
equilibrium solutions (by \textit{plus}) are: ($0^{\circ}$, 0.40)
and ($0^{\circ}$, 0.13), respectively. The semi-analytical contour
chart exhibits a good agreement with the numerical investigations
of maintaining the eccentricities with larger
$\theta_{3}$-libration amplitudes. On the other hand, the figure
implies  that the $\theta_{3}$-libration will be broken up if the
libration amplitudes exceed or approach the critical value of
$90^{\circ}$ near the separatrix, then two successive scenarios
may occur, either aligned orbits evolve into Type I geometry that
simply $\theta_{1}$ librates, or the system destabilize on the
shorter timescale. In the sense of the present solutions, the
aligned configuration for HD 82943 is not robust than the
antialigned case, where a wider relative periapse longitudes can
be initially adopted (see Figure 4). Bois (2003, private
communication) also confirmed that the stable strip of the
alignment for the initial $\varpi_{1}-\varpi_{2}$ is much narrower
than that of the antialignment in their numerical calculations for
HD 82943. However, if the aligned orbits are close to librating
center, the robustness can be strengthened with a different story.

In order to distinguish the dynamical behaviors in orbital
parameter spaces and explore how the stability zones dependence on
these parameters, we extensively perform abundant integrations for
each series (where in each series, we only allow one parameter
vary but others keep unchanged) to numerically search for the
stable solutions. In the following sections, we will discuss these
results respectively.

\subsubsection{Dependence on the semi-major axes}
Figure 7a show the numerical scanning of the HD 82943 system in
the semi-major axes space of [$a_{1},a_{1}$]. All other orbital
parameters are taken from Table 2. In this group of the
simulations, we choose $a_{1}\in$ [0.60 AU, 0.80 AU] and
$a_{2}\in$ [1.00 AU, 1.30 AU] with the resolution of $20 \times
20$ grids, each direction spaced 0.01 AU. In Fig. 7a, the cross
indicates the unstable orbits and the filled circles for stable
orbits that the planets survive at least $10^{6}$ yr. Here, we
notice that there exists a stable strip with the width from $\sim
0.01$ AU to $\sim 0.02$ AU, indicating that these stable solutions
are quite close to the 2:1 resonance. Meantime, the intersection
between the vertical and horizontal lines represents for the
present solution from Fit 2, which is well located in the stable
map.

In the case of 2:1 resonance, we have $a_{2}/a_{1} =
(P_{2}/P_{1})^{2/3}$, where $P_{2}/P_{1}\simeq 2$, if both of
orbits contract or expand $k$ times of the initial configuration
(e.g., Fit 2) and the planetary masses becomes $\sqrt k$ of the
initial masses (where other orbital parameters remained), the
semi-amplitude of the radial velocity will not change, that's the
reason why we can see a linear stable ribbon in Fig. 7a. In this
sense, the semi-major axes cannot be uniquely determined unless
their orbital periods are previously obtained from the
observations.

\subsubsection{Dependence on the eccentricities}
Mayor et al. (2004) pointed out that one should be careful of
employing the newly updated orbital fits (Fit 2) due to the
derived best-fit solutions of lack of considering the
planet-planet interaction. Hence, the exploration for the orbital
parameter space is extremely useful to help understand the
stability regions relying on the eccentricities. Similar to the
numerical scanning in the semi-major axes space, here only both of
the eccentricities of $e_{1}$ and $e_{2}$ are allowed to be free
parameters. In this series of the investigations, we let
$e_{1}\in$ (0.0, 0.60 ) and  $e_{2}\in$ (0.0, 0.60) and the
resolution in [$e_{1}$, $e_{2}$] space is $20 \times 20$ points,
spaced 0.03 in each grid (see Figure 7b), the integration time is
$10^{6}$ yr for each orbit. In the figure, the symbols are defined
as in Figure 7a, where the filled circles mean those orbits are
steady over secular evolution.

According to the eccentricity of the outer planet $e_{2}$, we
summarize the results as follows:

(1)strongly stable cases, where  $0 < e_{2}\le 0.24$ (40\%),

(2)partially stable cases, where  $0.24 < e_{2}\le 0.39$ (16\%
stable, 9\% unstable),

(3)unstable cases, where  $0.39 < e_{2} < 0.60$ (35\%).

As for case (1), we should mention two points: firstly, the
eccentricities of Fit 2 (shown by the crossing of two lines) just
reside in the stable map of [$e_{1}$, $e_{2}$]; secondly, in this
figure, we notice that there exist stable solutions when
$e_{1}=0.4$ and $e_{2}\simeq 0$, which is consistent with Mayor et
al. (2004), who also provided the other solution involved in an
aligned configuration with a near-circular orbit for outer planet
but found the residuals do not dramatically change. We can also
note that a higher $e_{1}$ can be possible in this system. Both of
cases (1) and (2) contain  more that 50\% of the stable orbits in
the numerical scanning, exhibiting that only a moderate $e_{2}$
will favor the stability of the HD 82943.

The case (3) shows that when  $e_{2}$ becomes larger than 0.39, no
regular orbits will appear even though $e_{1}$ can be adopted to
be small values, which suggests that in this situation the
stability for HD 82943 is more sensitive to the orbital motion of
the outer planet, which can greatly determine whether the system
is in chaotic or regular status.

\subsubsection{Dependence on the mean anomalies and arguments of periapse}
Figures 7c and 7d  exhibit our numerical exploration in the
parameter spaces of [$M_{1}, M_{2}$] and [$\omega_{1},
\omega_{2}$] for the time span of $10^{6}$ yr, respectively, with
the resolution of $18 \times 18$ grids. The stable map in the
[$M_{1}, M_{2}$] space again show the condition of the 2:1 MMR ,
where there is a linear relationship $M_{1}=2M_{2} + const$.
Therefore, on the basis of Fig. 7c, we can not only verify our
original mean anomalies of Fit 2 (by intersection) but find the
other arguments related to stable orbital motions for two planets
. The stable diagram in the [$\omega_{1}, \omega_{2}$] space
reveals the fact that both of the apsidal longitudes of
$\varpi_{1}$ and $\varpi_{2}$ (where nodal longitudes are constant
here) precess at the equal rate about the star. The stable zones
in Fig. 7d are quite narrower, as we stated before, this results
from the aligned configuration with a larger libration of
$\theta_{3}\sim 80^{\circ}$ (ref. Fig. 5).  Both figures clearly
illustrate that those stable configurations are linked to the 2:1
resonance and apsidal corotation.

\subsubsection{Dependence on the planetary mass ratio}
In this section, we expect to understand how the stability depends
on the planetary mass ratio for the coplanar planetary
configurations. Here in our numerical study, we still adopted the
orbital elements from Table 2, except the masses of two planets
$m_{1}$ and $m_{2}$. Firstly, we varied $\sin i$ in an incremental
step of 0.05 ranging from 0.35 to 1.00, and the rescaled masses of
$m_{k}$ ($k=1, 2$) can conveniently obtained by the product of
$m_{k} \sin i$ for a specific $\sin i$, where  the mass ratio
$m_{1}$/$m_{2}$ is fixed to be $1.85/1.84 \sim1$. Then 14 coplanar
integrations were carried out for $10^{7}$ yr in this runs.  The
numerical results show that the stable orbits (of Type II) for HD
82943 requires $\sin i \geq 0.50$. The stability set limitations
on the planetary masses where $m_{1}\in [1.85 M_{Jup}, 3.70
M_{Jup}]$ and $m_{2} \in [1.84 M_{Jup}, 3.68 M_{Jup}]$ for
different values of $\sin i$. In the case of GJ 876, similarly,
Marcy et al. (2001) reported that several stable systems,
catalogued in Type II, had $ \sin i = 0.50$ for the integration
timescale up to 500 Myr.  Additionally, Laughlin \& Chambers
(2001) also suggested that $\sin i$ possibly ranges from 0.5 to
0.8 for the coplanar fits for GJ 876. From above statements, we
notice that $\sin i \geq 0.50$ is necessary for both systems
occupying a 2:1 resonance and the extent of the planetary masses
can further be estimated.

In the other computations, we changed the mass ratio for
$m_{1}$/$m_{2}$ and fixed the mass $m_{2}=1.84$ $M_{jup}$, let
$m_{1}$/$m_{2}\in [1/4, 4]$, then performed over thirty additional
simulations to explore the stability of the system depending on
the variational mass ratios. As a result, we found that
$m_{1}$/$m_{2}\leq 2$  is  necessitated for the stable orbits,
indicating an upper limitation for the mass of the inner planet
when we remain that of the outer companion. In the following step,
we replaced $m_{1}$ with much lower masses ranging from several
$M_{\oplus}$ to 20 $M_{\oplus}$ for new investigations, and found
that the stable geometry of Type II can still be remained and a
surprise is that the eccentricity of the inner planet can amount
up to $\sim 0.90$ with quasi-periodic oscillations. These
numerical explorations suggest that the apsidal configuration can
maintain with much smaller masses for inner planet, which is
consistent with the results  given  by Beauge et al. (2003), who
analyzed the corotational solutions by gradually decreasing the
mass of the inner planet and found that although the solutions
simply depend weakly on the individual masses of the planets and
the maximum value for the planetary mass of $m_{1}$ is indeed in
existence with given eccentricities for the stable apsidal
topology.

\subsubsection{Dependence on the relative inclination}
In our solar system, most of the major planets have small
inclinations with regard to the essential reference plane.
However, there is a special case for the Neptune-Pluto system
($M_{Neptune} \approx \frac{1}{5} M_{Saturn}$), where Pluto orbit
an eccentric trajectory with a large inclination of $\sim
17^{\circ}$ but Neptune travels on a near-circular orbit with the
inclination less than $2^{\circ}$ (Murray \& Dermott 1999). Cohen
\& Hubbard (1965) discovered that these two planets are involved
in a 3:2 resonance with the argument
$3\lambda_{P}-2\lambda_{N}-\varpi_{P}$ ($\lambda_{P,N}$ denote,
respectively, the mean longitudes of Pluto and Neptune,
$\varpi_{P}$ is the longitude of perihelion of Pluto). These
studies reveal that there may exist a relative inclination
configuration between two orbits in association with the mean
motion resonance. In this sense, we anticipate to understand the
situation for the present exosystem.

We again performed extra tests of mutually inclined orbits for two
resonant planets for the integration of $10^7$ yr, to examine the
stability of the HD 82943 system. The orbital data were taken from
Table 2 and for simplicity, we just adopt their minimum masses and
remain them unchanged for our additional integrations. For the
initials, here we simply altered the inclination of the outer
planet $I_{2}$ and increased $I_{2}$ in an increment of
$5^{\circ}$ from $5.5^{\circ}$ to $85.5^{\circ}$, but always
assumed the inclination of the inner planet $I_{1}$ to be
$0.5^{\circ}$. The numerical integrations reveal the fact that
although the inclination was greatly changed, the two planets are
still in 2:1 MMR and apsidal alignment (Type II) with the
inclination of $I_{2}\leq 25.5^{\circ}$. On the contrary, for
$I_{2}>25.5^{\circ}$, we found that the eccentricity of the inner
planet can pump up to unity due to the great perturbation exerted
by the outer planet and  leave the orbits eventually. The relative
inclination $i_{r}$ can be determined by $\cos i_{r}=\cos I_{1}
\cos I_{2} + \sin I_{1} \sin I_{2} \cos (\Omega_{1} -\Omega_{2})
$, where $\Omega_{1}$ and $\Omega_{2}$ are, respectively, the
longitudes of ascending nodes for two planets. To sum up, in the
case of the non-coplanar configuration for HD 82943, stability
requires that the relative inclination be $\sim 25^{\circ}$ or
less. In the case of GJ 876, Rivera \& Lissauer (2001) carried out
dynamical fits for the possible best solutions for the mutual
inclination geometry and found that the stability for GJ 876
requires the relative inclination to be $12^{\circ}$ or less.
Therefore, it should be mentioned that the mutually inclined
orbits dynamically survive over long-term timescale in the
planetary geometry locking into a 2:1 MMR, but it should place
constraints on the initial conditions for two planets. Thommes \&
Lissauer (2003) showed that two planets migrating in resonance can
perform significant excitation of their inclinations and the
relative inclination between two planets can grow to $\sim
30^{\circ}$ in their three-dimensional simulations. As a final
note for this part, the above-mentioned numerical investigations
for various mutual inclined exosystems show that the low relative
inclination do play an important role in the stability for such
systems, and this mutual inclined configuration may arise from the
inclination-type resonance (Thommes \& Lissauer 2003) associated
with the longitudes of the ascending nodes for inner and outer
planets if the inner planet owns a high eccentricity (e.g., for
Fit 2, $e_{1}=0.38$), while the higher relative inclination
geometry can destabilize the system (e.g., Lissauer \&  Rivera
2001) on shorter timescales.

\subsection{Comparison with other works}
As we mentioned previously, several studies have been devoted to
explore the  orbital motions of the exosystems in 2:1 MMR. In the
case GJ 876, Lee \& Peale (2002) discovered that there are stable
configurations with $\theta_{1}$, $\theta_{2}$ and $\theta_{3}$
all librating about $0^{\circ}$ (related to Type II) for $0.15
\lesssim e_{1} \lesssim 0.86$, and Kinoshita \& Nakai (2001) and
Ji et al. (2002) also studied this system and confirmed the 2:1
resonance acting as an effective mechanism to hold the system,
moreover, an extended stable geometry of Type I was shown by Ji et
al. (2003b) for GJ 876. In other works, Hadjidemetriou \&
Psychoyos (2003) numerically studied the families of periodic
orbits for HD 82943 in the rotating frame, and indicated that two
kinds of families orbits can survive for the 2:1 resonant planets
in their computations. However, it is the first time that we show
an exhaustive exploration in the orbital parameter space for the
HD 82943 system based on the updated solutions, which reveals the
new dynamical features for this system.

In summary, the results presented in this article on the possible
stable planetary geometry involved in a 2:1 resonance are not only
consistent with the previous studies, but also outline an
extensive dynamics of these systems.  Moreover, we point out that
the aforementioned three groups of resonant configurations imply a
general principle that governs two planets in a dynamically stable
state over secular orbital evolution in the planetary systems,
which should be examined by more future discoveries of the 2:1
resonant pairs. Still, we will concentrate on whether a postulated
terrestrial can survive in such geometry (see \S4), and how a
certain system is evolved into an apsidal alignment or
antialignment resonant configuration in \S5.

\section{Habitable zones}
The Habitable zones (HZ)  are generally convinced to be suitable
places for terrestrial planets that can provide the liquid-water,
subtle temperature and  atmosphere environment, and other proper
conditions (Kasting et al. 1993), supporting the development and
biological evolution of life on their surfaces. The habitable
region is considered to be centered at $a_{p}$ = 1 AU
$(L/L_{\odot})^{1/2}$ (Gould, Ford, \& Fischer 2003) or
$a_{p}\geq$ 1.03 AU $(L/L_{\odot})^{1/2}$ (Kuchner 2003), such
that a water planet can exist in equilibrium with stellar
radiation, where $a_{p}$ is the radius of the planet's orbit and
$L$ is the luminosity of the host star. In a viewpoint of the
dynamics, first of all, there should be stable orbits at $\sim 1$
AU  for Earth-like planets moving about main sequence stars over
very long timescale and we can also refer this to the dynamical
habitability. In a recent work, Menou \& Tabachnik (2003)
exhaustively studied the dynamical habitability of 85 known
extrasolar planetary systems through numerical simulations of
their orbital dynamics in the presence of potentially habitable
terrestrial planets. They found that more than half of the known
exosystems cannot harbor habitable terrestrial planets and about
25\% of the systems mostly with close-in giant planets are
dynamically habitable, similar to our solar system. And in this
section, we only draw our attention to the potential habitable
zones for the systems (e.g. HD 82943 and GJ 876) with a stable
geometry in a 2:1 resonance.

Here we performed numerical surveys in examining the Habitable
zones both for HD 82943 and GJ 876. And  at first, we considered
stable configurations for HD 82943 (Fit 2) and GJ 876 (Keck and
Lick Fit with $\sin i=0.78$; see Table 2 in Laughlin \& Chambers
2001), which can be viewed as genuine systems closest to the
actual observations. Then, we generated 100 seed planets that all
bear the same masses of Earth (where $M_{\oplus} \simeq 3.14
\times 10^{-3} M_{jup}$) in the simulations for each system. In
Figure 8, we can see the distribution of the initial orbital
elements for these postulated planets--they initially move on the
less inclined belt with the relative inclination with regard to
the plane of the resonant pair no more than $5^{\circ}$, the
semi-major axis $a\in$ [0.96 AU, 1.05 AU], the eccentricity
$e\in[0,0.1]$, and the leftover arguments are set at random. Here
we simply accounted for the near-circular orbits rather than much
more eccentric orbits, because the life should develop and evolve
in the biological environment without larger variations of
temperature, implying the Earth-like planet should not move
neither too close nor too far from the parent star.

For each star-two-planet-"Earth" four-body system, we carried out
the integration for the timescale of $10^{7}$ yr, to examine the
dynamical habitability for the HD 82943 system. As we shall see,
we find that none of the orbits can retained stable in the region
at $\sim 1$ AU  for the integration time, the seed planet travels
on hyperbolic or parabolic orbits at the characteristic timescale
$ \tau < 2 \times 10^{5}$ yr (where 81\% of them are ejected at $
\tau \le 5 \times 10^{4}$ yr and 95\% at $ \tau \le 1.0 \times
10^{5}$ yr, see Figure 9a), owing to the scattering (Lin \& Ida
1997) between two close resonant planets for the excitation in the
eccentricity. A typical orbital evolution for the ejected orbits
is shown in Figure 9b, the semi-major axe $a$ grows from $\sim 1$
AU to over 100 AU, the eccentricity $e$ undergoes a rapid increase
from $\sim 0.1$ to 1 and the inclination $i$ is also excited to a
high value of $\sim 40^{\circ}$ (in some cases, the inclination
can exceed $90^{\circ}$), finally, at $4,800$ yr, the assumed
Earth-like planet is scattered away from the two-planet system,
while the orbital motions of the resonant planets remain as usual.
The gravitational scattering on the terrestrial  planet caused an
unstable motion either as an ejection or a catastrophic collision
with the star or planets due to the dynamical instabilities (Ford,
Havlickova, \& Rasio 2001). Our numerical investigations suggest
that there will be least possibility for stable Earth-like orbits
surviving in the range of [0.73 AU, 1.18 AU], even for wider
habitable regions [0.75 AU, 1.40 AU] given  by  Kasting et al.
(1993), which is in good agreement with the simulations by Menou
\& Tabachnik (2003). Analytically,  Gladman (1993) attained the
minimum separation $\Delta$ between two planets moving on
initially circular orbits require for stability,
\begin{equation}
\label{16} \Delta = 2\sqrt 3 R_{H},
\end{equation}
where the mutual Hill radius  $R_{H}$ is
\begin{equation}
\label{17} R_{H}={\left(\frac{m_{1} + m_{2}}{3m_{0}}\right)}^{1/3}
\left({\frac{a_{1} + a_{2}}{2}}\right).
\end{equation}
For Fit 2, we achieve $\Delta \simeq 0.34$ AU and $R_{H} \simeq
0.10$ AU, which these values may become a bit different as the two
planets for HD 82943 orbit on eccentric paths. Nevertheless, our
goal here is to present a qualitative analysis, and we can easily
notice that the initial orbits for the assumed planets are quite
close to Hill sphere regime, resulting in their unstable motions.
But it cannot rule out other stable regions for additional
companions. For example, Sandquist et al. (2002) explored a
three-planet case for HD 82943, which a supposed jovian planet
with the mass of 0.5 $M_{jup}$ at 0.02 AU was added to the system,
and they found that this additional planet does not affect the
orbital motion of the two outer planets and also not disturb the
stability of this system. On the other hand, for the exterior
domain extending to $30$ AU $\sim 50$ AU or even farther out, does
there exist the dust rings or circumstellar disk that are
analogous to the Kuiper Belt  being a remnant of the solar nebula?
The recently launched SIRTF mission will provide fruitful
information on this point in its ongoing tasks.

As a comparison, we accomplished the other same runs for GJ 876.
And we conducted the four-body integrations for the time span of
integration of $10^{6}$ yr ($\sim 10^{7}$ orbital periods for the
innermost planet) and found all the systems are stable for the
integration timescale. Figure 10 exhibits the typical orbital
evolution for the Earth-like planet, both the semi-major axis $a$
and the eccentricity $e$ execute small fluctuations about 1 AU and
0.06, respectively, and the inclination $i$ also remains less than
2 degrees in the same time span. And there is no sign to indicate
that such regular orbits at $\sim 1$ AU with low eccentricities
will become chaotic for much longer time even for the age of the
star. Due to the faint luminosity for M4 star GJ 876, the
habitable zones may reside in the range of [0.10 AU, 0.20 AU]
(Kasting et al. 1993; Menou \& Tabachnik 2003) sufficiently nearer
to the orbits of GJ 876 b ($\sim$ 0.13 AU) and GJ 876 c ($\sim$
0.21 AU) (Marcy et al. 2001; Laughlin \& Chambers 2001; Rivera \&
Lissauer 2001), hence, the global chaos, corresponding to the
circumstance of a collisional disruption or ejection, can always
take place according to the Hill stability criterion (Gladman
1993; Chambers, Wetherill \& Boss 1996) or direct numerical
examinations, and exclude the possible existence of Earth-like
planets for such habitable regions. Whereas Kuchner (2003) pointed
out that a liquid planet with $a_{p}\geq$ 1.03 AU
$(L/L_{\odot})^{1/2}$ (where for GJ 876, $a_{p}\geq 0.12$ AU) may
exist under the protection of a thick steam atmosphere, it seems
plausible for the Earth-like planets revolve around GJ 876 at the
orbit about 1 AU, in the concept of the dynamical habitability.
However, our predictions  should be confirmed through careful
investigation in future research if the Earth-like planets do
survive secularly in such habitable zone. Firstly, the direct
evidence of the presence for such planets strongly depend upon the
future high resolution astrometric measurements , e.g., Space
Interferometer Mission (SIM) or Terrestrial Planet Finder (TPF)
that can provide much better precision than the present long-term
precision of $\sim$ 2-3 m s$^{-1}$ (Butler et al. 2003) for
ground-based Doppler surveys.  Secondly, the simulations of the
runaway accretion of the planetary embryos (Laughlin et al. 2002)
suggest that it is impossible for the Earth-mass planets to be
formed in Habitable zones even for a well-separated system with
near-circular orbits (e.g., 47 Uma), because the formation of the
terrestrial planets are terrifically constrained by giant planet
migration (Armitage 2003), therefore the depletion of small
planetesimals in a disk-clearing phase may possibly take place
when the massive planets sweep over them in their orbital decay
process unless this orbital migration can halt at a proper
position allowing for the continuous growth of the planetary
embryos or the already-created Earth-like planets can also follow
the migration with the giant planets in step and cease about the
Habitable regions. Nevertheless, fortunately, the terrestrial
planets are conceivable to survive in a modest fraction of systems
where a single generation of massive planets formed (Armitage
2003) without significant migration.

\section{The Possible Origin for the Orbital Geometry}
The short-period giant planets that are close to the parent star
with high temperature (e.g., 51 Peg, so-called hot Jupiter with
$a<0.1$ AU), are now thought to be formed through the shrinking
orbital migration rather than be formed \textit{in situ} due to
the tidal interaction of a protoplanet with a surrounding gaseous
disk (Ward 1997) and this orbital migration can originate from the
exchange of angular momentum between the natal disk and
protoplanet (Papaloizou 2003 and references therein). Now, it is
widely believed that the migration will ubiquitously come about
when two-planet systems suffer tidal interaction with an exterior
disk, the outer planet will migrate inward until it gets trapped
in a low order mean motion resonance with the inner companion
(Bryden et al. 2000; Kley 2000) and the eccentricities will stop
growing and be balanced as steady equilibrium values (Nelson \&
Papaloizou 2002) after being captured into the resonances.

For the aligned orbits of Type II, in the earlier works for GJ
876, Snellgrove et al. (2001) performed full hydrodynamical
two-dimensional simulations to study the inward migration for GJ
876 planets interacting with each other and a protoplanetary disk,
and the outer planet can be trapped into a 2:1 commensurability
with the inner planet through slow migration, where the orbits
were observed to be aligned with the resonant angles librating
about zero (see also an overview in Papaloizou 2003) and the
system can survive at least $2\times 10^{7}$ orbits by the removal
of the external disk. Subsequently, Lee \& Peale (2002) showed
that the forced inward migration of the outer companion of the GJ
876 system leads to certain capture into the observed resonances
if the initial eccentricities are smaller enough where $e_{1}\leq
0.06$ and $e_{2}\leq 0.03$ with proper migration rates. They also
revealed that after resonance capture, the eccentricities increase
rapidly due to resonant interactions between the planets and the
forced migration without eccentricity damping, while with
parameterized eccentricity damping where $\dot e_{i}/e_{i}=-K|\dot
a_{i}/a_{i}|$, the eccentricities can reach equilibrium values
that remain unchanged for sufficiently long migration in the
resonances. Recently, in the study of the evolution in resonant
planetary systems (e.g., HD 82943), Kley et al. (2004) found that
the two planets can enter into a 2:1 resonance in varying
timescale provided that $e_{2}<0.25$, with both full
hydrodynamical simulations and damped N-body calculations. They
further suggested that there are several requirements for the 2:1
resonance capture--the more massive mass for the outer companion,
the higher eccentricity of the inner planet and the apsidal
alignment of $\theta_{3}=0^{\circ}$. In the case of Type II (Fit
2), where the eccentricity $e_{2}=0.18$ and a moderate
$e_{1}=0.38$ and with  two nearly equal planetary masses, or in
the case of Fit 1b, where there can also be stable geometry about
$e_{2}\sim 0.25$ and a higher $e_{1}$ with a larger outer planet,
in both cases the two planets undergo $\theta_{3}$-libration about
$0^{\circ}$, thus we may comment that if the HD 82943 system is
most likely to be aligned (Lee M.H. 2003, private communication),
then the origin for such orbital geometry can be possibly attained
by the predictions of the hydrodynamic models (Kley et al. 2004),
we will address this problem in our forthcoming paper.

For the antialigned orbits of Type III, Thommes \& Lissauer (2003)
argued that if the planets reached this configuration by migration
in resonance, a significant mutual inclination between their
orbits would seem to be required. As in this geometry, the two
orbits for HD 82943 can experience the intersection on the
co-orbital region (Ji et al. 2003c; Lee 2004), therefore a
moderate inclination between two orbits can be achieved to act as
a protection mechanism against close approaches (Thommes \&
Lissauer 2003). In \S3.3.2, we also show that the highly inclined
resonant configuration for this system can also be in presence,
which the system is stabilized by both the 2:1 resonance and the
apsidal phase-locking. However, for Fit 1a, although  the HD 82943
system can be stable at least for $10^{7}$ yr, it is still
difficult to secure the resonant capture with N-body models due to
the larger eccentricities of 0.41-0.54 in Kley et al. (2004), the
origin for such configuration may attribute to an additional
passing star's influence or pure gravitational scattering
mechanism (Lin \& Ida 1997; Adams \& Laughlin 2003), to pump up
the eccentricities. One of the final fates in the scattering
procedure is that the inner planet is left alone in the system
with an eccentric orbit while the outer planet is ejected (Ford et
al. 2001) into the stellar space, nearly immune to the
gravitational influence from the parent star. If the scattering
does happen in the course of the migration, a casual rapid capture
may be required to prevent one of the planets from throwing away
from the system and subsequently two orbits are well locked into
2:1 resonance (see \S3.2.1) at larger eccentricities.
Alternatively, another possible scenario is that both of the
eccentricities are excited to moderate values when the 2:1
resonance is being crossed (Chiang et al. 2002), subsequently the
continuous resonance crossings may terminate for the suitable
eccentricities evolved into the resonance geometry.

\section{Summary and Remarks}
In this paper, we have explored the stable geometry for a system
with two planets involved in a 2:1 MMR, and we mainly concentrate
on the study of the HD 82943 system by adopting two different sets
of the orbital parameters. In our numerical simulations, we found
there are three possible stable configurations for HD 82943
system:(1)Type I, only $\theta_{1} \approx 0^{\circ}$, (2)Type II,
$\theta_{1}\approx\theta_{2}\approx\theta_{3}\approx 0^{\circ}$
(aligned case), and (3)Type III, $\theta_{1}\approx 180^{\circ}$,
$\theta_{2}\approx0^{\circ}$, $\theta_{3}\approx180^{\circ}$
(antialigned case). The direct integrations show that all stable
orbits are related to the 2:1 commensurability that remains the
semi-major axes for two companions slightly vary about the initial
values over the secular dynamical evolution for $10^{7}$ yr,
further the apsidal phase-locking for two planets can enhance the
stability for this system, because the eccentricities are
simultaneously preserved to prevent the planets from frequent
close encounters. Additionally, we comment that other systems
trapped in 2:1 MMR (e.g. GJ 876 or possibly HD 160691) can select
one of the aforementioned three stable topology in their authentic
orbital motions, which are determined by their initial parameters.
And we underline that such configurations, making up the dynamical
families, can serve as a general regulation for the long-term
stable orbits of the 2:1 resonant exosystems. In final, we
summarize some conclusions:

Using the earlier best-solutions of Fit 1, we found three types of
stable orbits for HD 82943, while we observed the steady
configurations of Types I and II by employing the new solution of
Fit 2 supplied by Mayor et al. (2004). In the meantime, we also
proposed a semi-analytical model to study the $e_{i}-\Delta\varpi$
Hamiltonian contour determined by the initial parameters, and  the
closest level curves encompass the direct numerical results, which
presents a good agreement between them. The theoretical figures
can still provide valuable information on the dynamics for two
planets.

Given the stable geometry based on Fit 2, we then extensively
examined the dependence of the stability of HD 82943 in the
orbital parameter space and the planetary mass ratios. In the case
of the non-coplanar circumstances, we found that stability
requires that the relative inclination be $\sim 25^{\circ}$ or
less. For a fixed planetary mass ratio $\sim 1$, the stable orbits
for HD 82943 requires $\sin i \geq 0.50$. For the non-constant
mass ratio case (where $m_{2}$ is kept), the requirement of the
stability is $m_{1}$/$m_{2}\leq 2$, which indicates an upper
boundary for the mass of the inner planet. Concerning the
eccentricities, the system can be always steady when $0 < e_{2}\le
0.24$ and $0 < e_{1}< 0.60$. In a word, these outcomes do
demonstrate that the stability for HD 82943 is strongly sensitive
to the sound planetary masses, rather lower relative inclination
between two orbits, the eccentricities and other orbital
parameters. In addition, we showed that the assumed terrestrial
bodies cannot exist in the habitable zones of HD 82943 thanks to
strong perturbations induced by two massive planets, but the
Earth-like planets can be dynamically habitable in the GJ 876
system at $\sim 1$ AU in the numerical surveys.

However, as the additional measurements will improve the present
best-fit solution for HD 82943 with a corrected model by taking
into account the planet-planet interactions. Thus, we hope to have
more precise orbital solution not only to check for our presented
predictions in this paper but to better make sense of the complete
dynamics and the origin of this system in the near future.

\acknowledgments {We thank G. Laughlin, M.H. Lee, S.J. Peale,
J.C.B. Papaloizou and E. Thommes for informative discussions and
helpful explanations, K. Gozdziewski for good comments. We are
grateful to M. Mayor and S. Udry for providing us their
manuscript. We also thank E. Bois for sending us his unpublished
figures. This work is financially supported by National Natural
Science Foundation of China (Grants No.10203005, 10173006,
10233020) and Foundation of Minor Planets of Purple Mountain
Observatory. This research has made use of NASA's Astrophysics
Data System (ADS).}

\clearpage

\appendix
\section{The form of the total angular momentum with use of Jacobi
coordinates}
The Hamiltonian with use of Jacobi coordinates has
the following form:
\begin{equation}
F=\frac{1}{2}\sigma_1\dot{\bf r_1}^2+\frac{1}{2}\sigma_2\dot{\bf
r_2}^2-\frac{Gm_0m_1}{r_{01}}
-\frac{Gm_0m_2}{r_{02}}-\frac{Gm_1m_2}{r_{12}},
\end{equation}
where
\begin{equation}
\sigma_1=\frac{m_0m_1}{m_0+m_1},\sigma_2=\frac{m_2(m_0+m_1)}{m_0+m_1+m_2}.
\end{equation}
Now we have two choices for the unperturbed Hamiltonian:\\
A)
\begin{equation}
F_0=\frac{1}{2}\sigma_1\dot{\bf
r_1}^2-\frac{Gm_0m_1}{r_1}+\frac{1}{2}\sigma_2\dot{\bf r_2}^2
-\frac{Gm_0m_2}{r_2}
\end{equation}
B)
\begin{equation}
F_0=\frac{1}{2}\sigma_1\dot{\bf r_1}^2-\frac{Gm_0m_1}{r_1}
+\frac{1}{2}\sigma_2\dot{\bf r_2}^2-\frac{Gm_2(m_0+m_1)}{r_2}.
\end{equation}
The difference between A) and B) is the last term. The choice A)
is recently popular. In the treatment of the triple star system
the choice B) is always adopted, since the form of the disturbing
function becomes simpler for the case $r_1<r_2$.

The unperturbed Hamiltonian takes the following form for each case:\\
A)
\begin{equation}
F_0=-\frac{Gm_0m_1}{2a_1}-\frac{Gm_0m_2}{2a_2},
\end{equation}
B)
\begin{equation}
F_0=-\frac{Gm_0m_1}{2a_1}-\frac{G(m_0+m_1)m_2}{2a_2}.
\end{equation}

The form of the total angular momentum $J$ is\\
A)
\begin{equation}
\sigma_1\sqrt{G(m_0+m_1)a_1(1-e_1^2)}+\sigma_2\sqrt{Gm_0\frac{m_0+m_1+m_2}{m_0+m_1}
a_2(1-e_2^2)},
\end{equation}
B)
\begin{equation}
\sigma_1\sqrt{G(m_0+m_1)a_1(1-e_1^2)}+\sigma_2\sqrt{G(m_0+m_1+m_2)
a_2(1-e_2^2)}.
\end{equation}
In the above expressions the meaning of the osculating elements
$a_2,e_2$ is different between A) and B), because the unperturbed
form is different. However, for the averaged Hamiltonian after the
elimination of the short periodic terms, the difference due to the
choice between the Jacobi coordinates and heliocentric (or
astrocentric) coordinates is the order of the perturbation
($m_1/m_0,m_2/m_0$), which could be negligible.

\clearpage

\begin{deluxetable}{lccccc}
\tablewidth{0pt} \tablecaption{The orbital parameters of HD 82943
planetary system (Fit 1\tablenotemark{a}).  $m_{0} = 1.05
M_{\odot}$} \tablehead{\colhead{} &\multicolumn{2}{c}{Fit 1a}
&\colhead{}
&\multicolumn{2}{c}{Fit 1b}\\
\cline{2-3}  \cline{5-6}\\
\colhead{Parameter} &\colhead{Outer (b)} &\colhead{Inner (c)}
&\colhead{} &\colhead{Outer (b)} &\colhead{Inner (c)}}

\startdata
$m$ sin$i$($M_{Jup}$)      & 1.63   & 0.88   & & 1.63   & 0.88   \\
$P$(days)                  & 444.6  & 221.6  & & 444.6  & 221.6  \\
$a$(AU)                    & 1.16   & 0.73   & & 1.16   & 0.73   \\
$e$                        & 0.41   & 0.54   & & 0.41   & 0.54   \\
$\Omega$(deg)              & 14.79  & 154.23 & & 327.65 & 239.11 \\
$\omega$(deg)              & 96.0   & 138.0  & & 96.0   & 138.0  \\
Mean Anomaly(deg)          & 97.86  & 10.02  & & 53.40  & 188.50

\enddata
\tablenotetext{a}{The orbital parameters for the planetary masses,
orbital periods, semi-major axes, eccentricities and apsidal
arguments are taken from the Geneva website, as of July 31, 2002.
Both of the inclinations are assumed to be $0.5^{\circ}$. Note Fit
1a and Fit 1b are, respective, one of Types III and II, amongst
100 first runs.}
\end{deluxetable}

\begin{deluxetable}{lcc}
\tablewidth{0pt} \tablecaption{The orbital parameters of HD 82943
planetary system (Fit 2\tablenotemark{a}).  $m_{0} = 1.15
M_{\odot}$} \tablehead{\colhead{Parameter} & \colhead{Outer (b)} &
\colhead{Inner (c)}} \startdata
$m$ sin$i$($M_{Jup}$)      & 1.84   & 1.85      \\
$P$(days)                  & 435.1  & 219.4     \\
$a$(AU)                    & 1.18   & 0.75      \\
$e$                        & 0.18   & 0.38      \\
$\Omega$(deg)              & 120.91 & 315.60    \\
$\omega$(deg)              & 237.0  & 124.0     \\
Mean Anomaly(deg)          & 168.56 & 185.12
\enddata
\tablenotetext{a}{The orbital parameters for the planetary masses,
orbital periods, semi-major axes, eccentricities and the
periastron arguments are adopted from Mayor et al. (2004), as of
October 13, 2003. The inclinations are the same as given in Table
1. The mean anomalies are derived at the epoch JD 2,452,396.82.
Fit 2 is one of Type II of 100 second runs.}
\end{deluxetable}
\clearpage

\begin{figure}
\figurenum{1} \plottwo{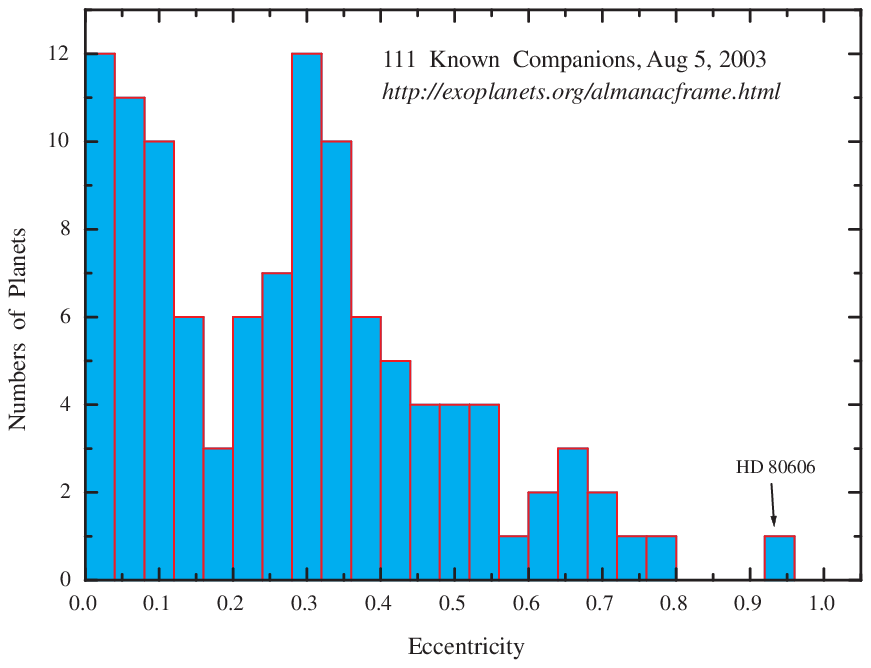}{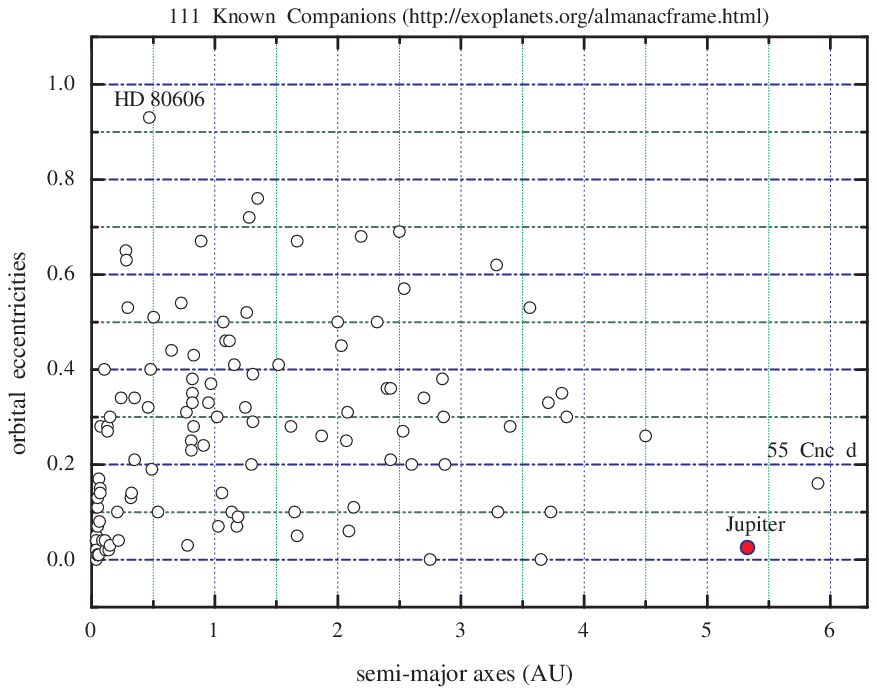} \caption{\textit{Left
panel}: the histogram for the eccentricities for 111 extrasolar
planets, as of Aug. 5, 2003. Notice that over 50\% of the planets
have the eccentricities larger than 0.30, and HD 80606 b can
occupy the eccentricity up to 0.93. \textit{Right panel}:
distribution of the semi-major axes plotted against the
eccentricities.  Note that 55 Cnc d can be distant as far as
nearly 6 AU (see also Marcy et al. 2003) from its parent star.
\label{fig1}}
\end{figure}
\clearpage

\begin{figure}
\figurenum{2} \plotone{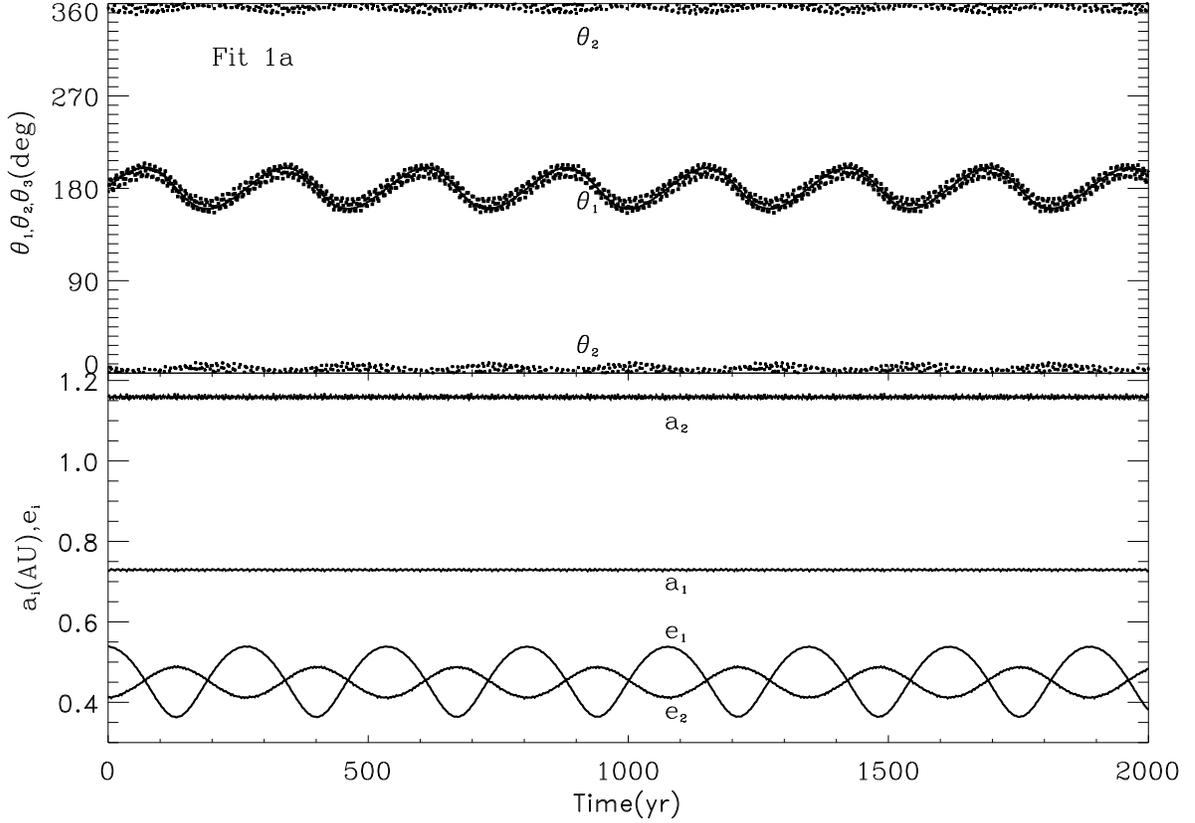} \caption{Orbital evolution for Fit
1a (antialigned orbits). \textit{Upper panel}: $\theta_{1}$
librates (by \textit{dots}) about $180^{\circ}$ with a moderate
amplitude of $\sim 30^{\circ}$, $\theta_{2}$ (by \textit{dots})
about $0^{\circ}$ with a small amplitude of $\sim 10^{\circ}$, and
$\theta_{3}$ (by \textit{thick line}) about $180^{\circ}$ with a
relatively small amplitude of $\sim 30^{\circ}$ for $t=10^{7}$ yr
(to see more clearly, we simply display a snapshot for $t=2000$
yr). \textit{Lower panel}: $a_{1}$ and $a_{2}$ slightly vibrate
about 0.73 and 1.15 AU for $t=10^{7}$ yr. Notice that $e_{1}$ and
$e_{2}$ separately reside in (0.36, 0.56) and (0.39, 0.49). The
2:1 resonance is confirmed by the modulations of the semi-major
axes $a_{i}$ and $\theta_{i}$ ($i=1,2$). Note that $\theta_{1}$
and $\theta_{3}$ share the same libration period of $\sim$ 300 yr.
\label{fig2}}
\end{figure}
\clearpage

\begin{figure}
\figurenum{3} \plotone{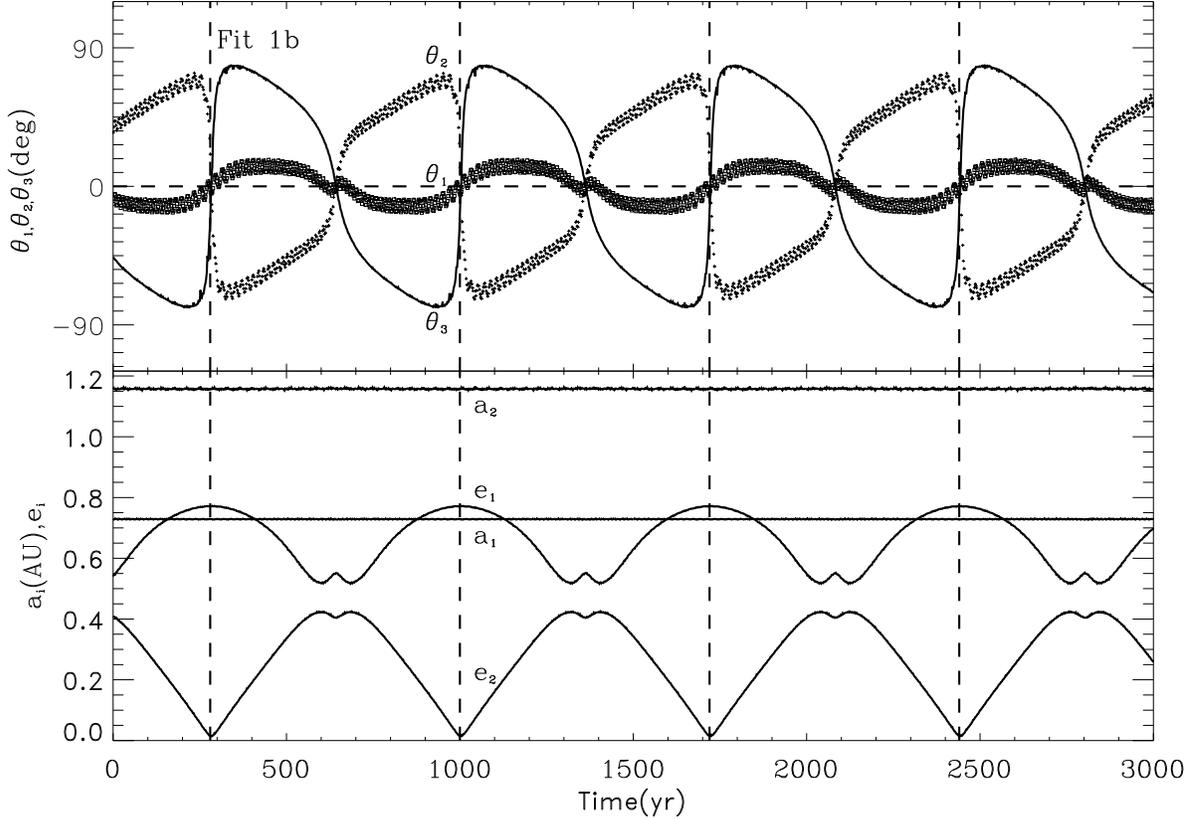} \caption{Orbital evolution for Fit
1b (aligned orbits). \textit{Upper panel}: $\theta_{1}$ (by
\textit{square sign}) librates about $0^{\circ}$ with a small
amplitude of $\sim 20^{\circ}$, but $\theta_{2}$ (by \textit{plus
sign}) and $\theta_{3}$ (by \textit{thick line}) individually
librate about $0^{\circ}$ (a snapshot for $t=3000$ yr ) with a
large amplitude of $\sim 70^{\circ}$. In this resonant aligned
geometry, we again observe that $\theta_{1}$, $\theta_{2}$ and
$\theta_{3}$ share the common librating periods of $\sim 700$ yr,
coupled with those of the eccentricities of $e_{1}$ and $e_{2}$.
\textit{Lower panel}: $a_{1}$ and $a_{2}$ do not change
dramatically but undergo small oscillations about 0.73 AU and 1.15
AU for $t=10^{7}$ yr, further the eccentricities $e_{1}$ and
$e_{2}$ librate in the range of (0.5, 0.8) and (0, 0.45),
respectively, in the behavior of the converse cycles, owing to the
conservation of the total angular momenta for the system.
\label{fig3}}
\end{figure}
\clearpage

\begin{figure}
\figurenum{4} \plottwo{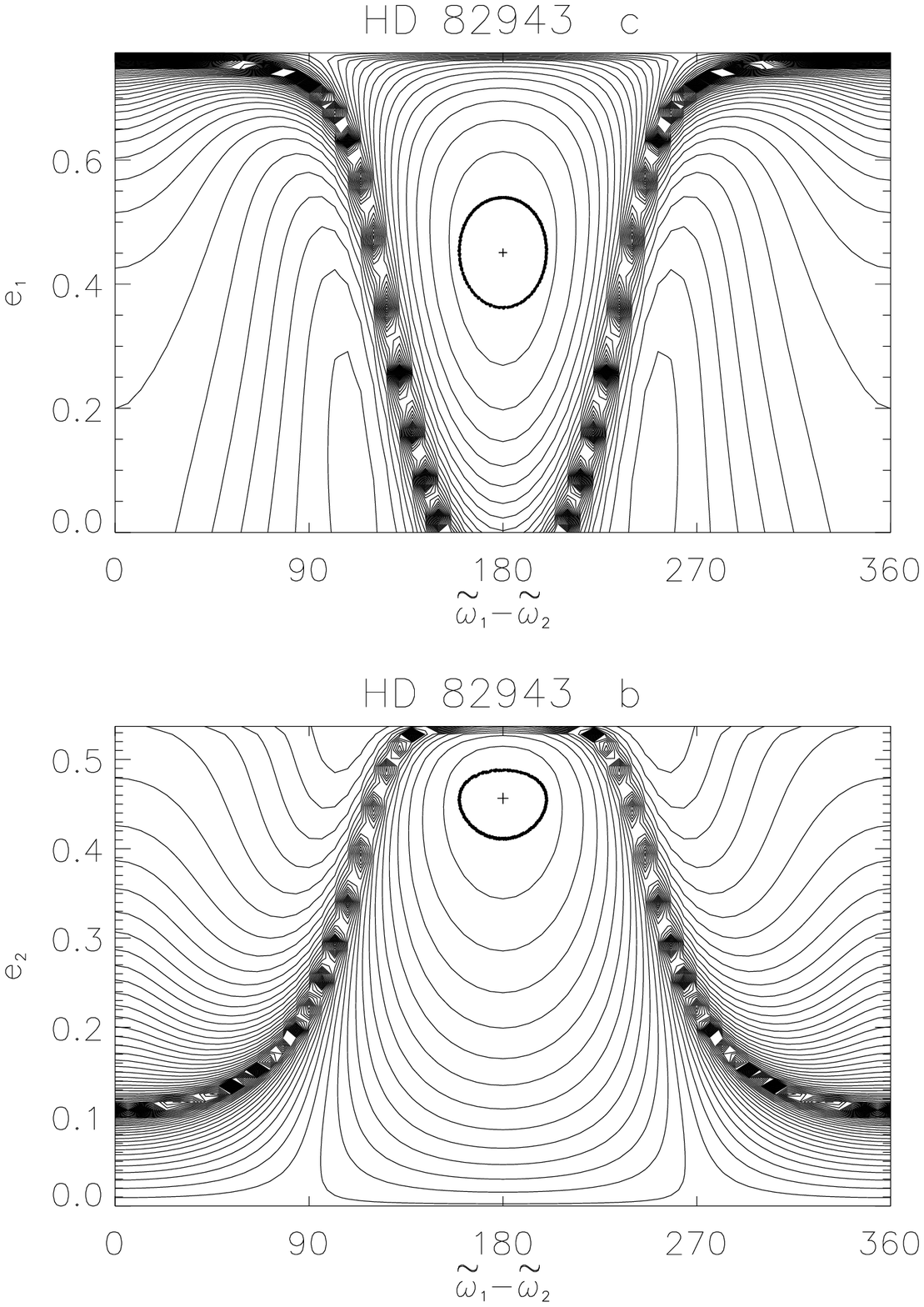}{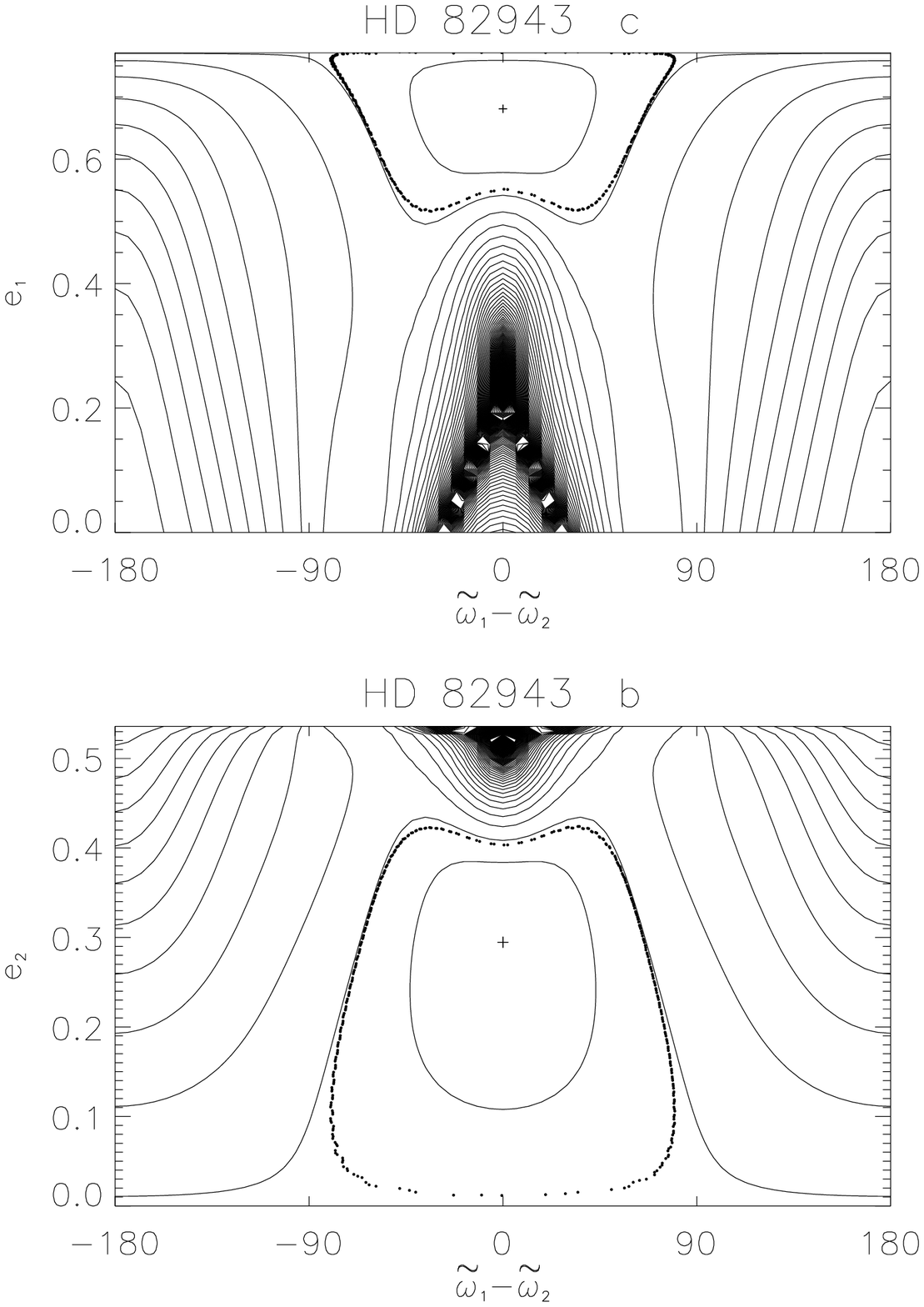}
\caption{$e$-$\Delta\varpi$ Hamiltonian contour map for Fit 1.
\textit{Left panel}: (a)For Fit 1a, the contour levels are
exhibited by \textit{thin line} and the numerical solutions shown
by \textit{thick line} (the innermost curve about plus). The
marked \textit{plus} denotes the stationary solutions --
($180^{\circ}$, 0.45) and ($180^{\circ}$, 0.46), respectively.
\textit{Right panel}: (b)For Fit 1b, \textit{thin line} for
contour levels and \textit{dotted points} for numerical solutions,
the librating centers are, respectively, ($0^{\circ}$, 0.68) and
($0^{\circ}$, 0.29). The $e_{i}-\Delta\varpi$ figures both show a
good agreement between the numerical results and the
semi-analytical solutions. \label{fig4}}
\end{figure}
\clearpage

\begin{figure}
\figurenum{5} \plotone{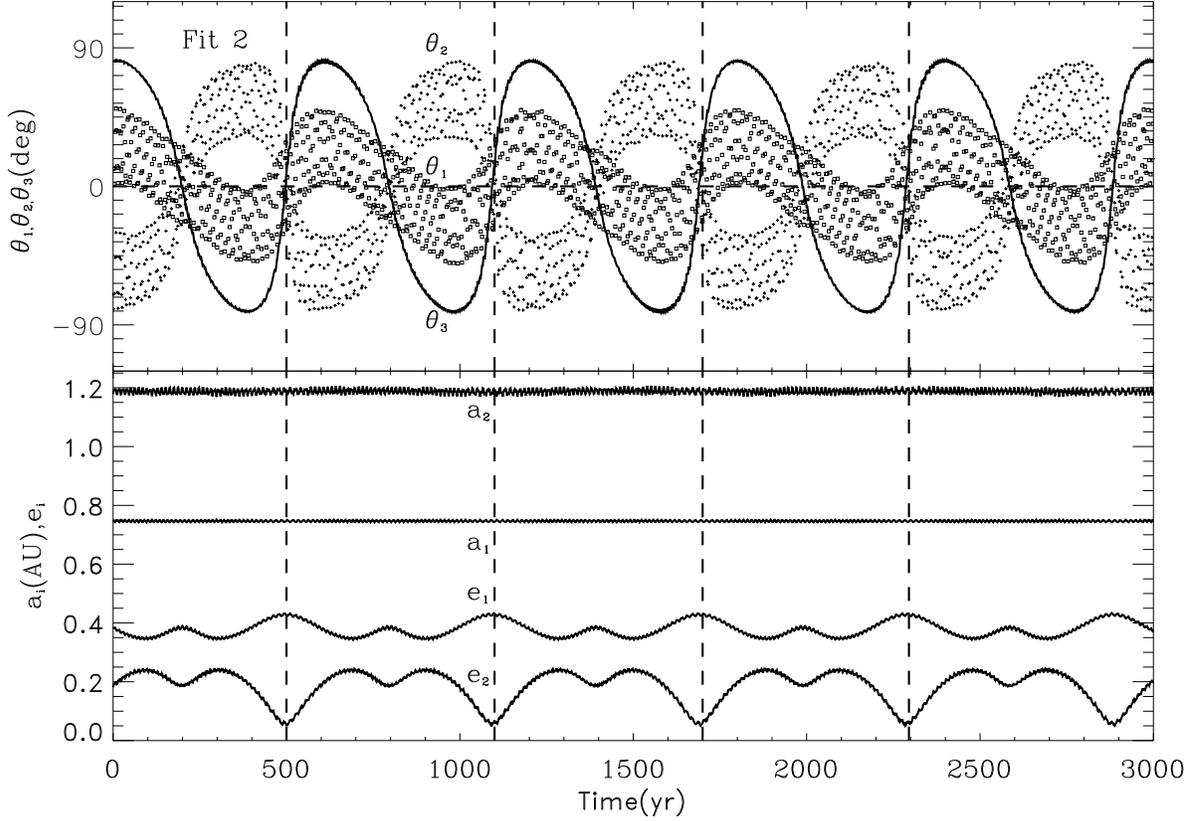} \caption{Orbital variations for Fit
2 (aligned orbits). \textit{Upper panel}: $\theta_{1}$ (by
\textit{square sign}) librates about $0^{\circ}$ with a moderate
amplitude of $\sim 45^{\circ}$, but $\theta_{2}$ (by  \textit{plus
sign}) and $\theta_{3}$ (by \textit{thick line}) individually
librate about $0^{\circ}$ (a snapshot for $t=3000$ yr ) with large
amplitude of $\sim 80^{\circ}$. \textit{Lower panel}: the
semi-major axes $a_{1}$ and $a_{2}$ are almost unchanged and they
modulate about 0.75 AU and 1.18 AU with relatively smaller
amplitude for $t=10^{7}$ yr, at this time, the amplitude of the
oscillations for $e_{1}$ and $e_{2}$ are not so large and they are
just wandering in the span (0.34, 0.44) and (0, 0.25),
respectively. Notice that the eccentricities have the libration
periods of $\sim$ 600 yr, which are coupled with those of
$\theta_{1}$ and $\theta_{3}$. \label{fig5}}
\end{figure}
\clearpage

\epsscale{0.6}
\begin{figure}
\figurenum{6} \plotone{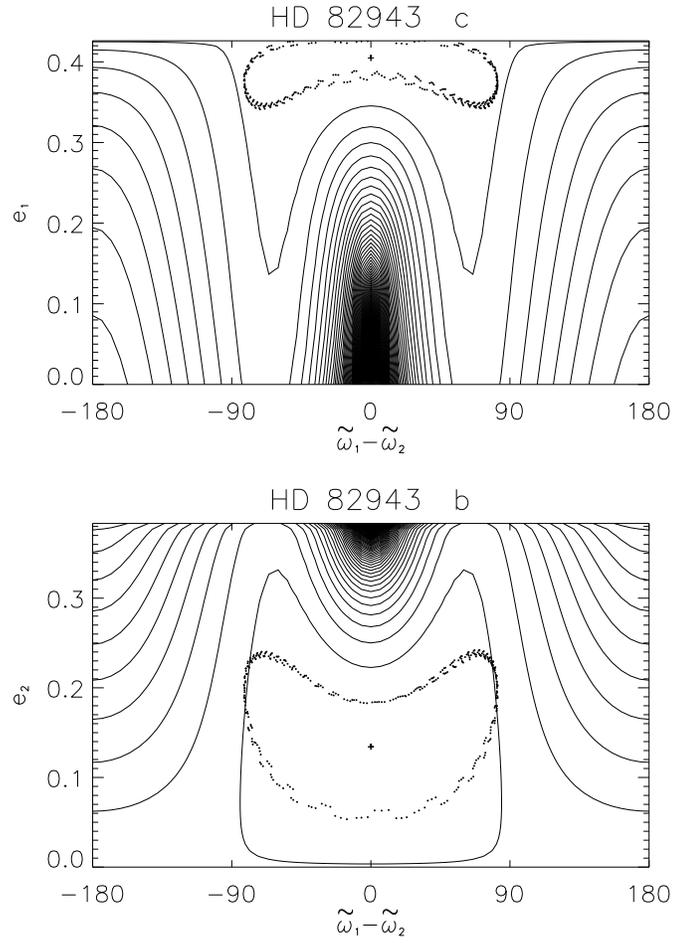} \caption{$e$-$\Delta\varpi$
Hamiltonian contour map for Fit 2. Same as the definition in
Figure 4b. The equilibrium points are: ($0^{\circ}$, 0.40) and
($0^{\circ}$, 0.13), respectively. \label{fig6}}
\end{figure}
\clearpage

\epsscale{1.0}
\begin{figure}
\figurenum{7a-7b} \plottwo{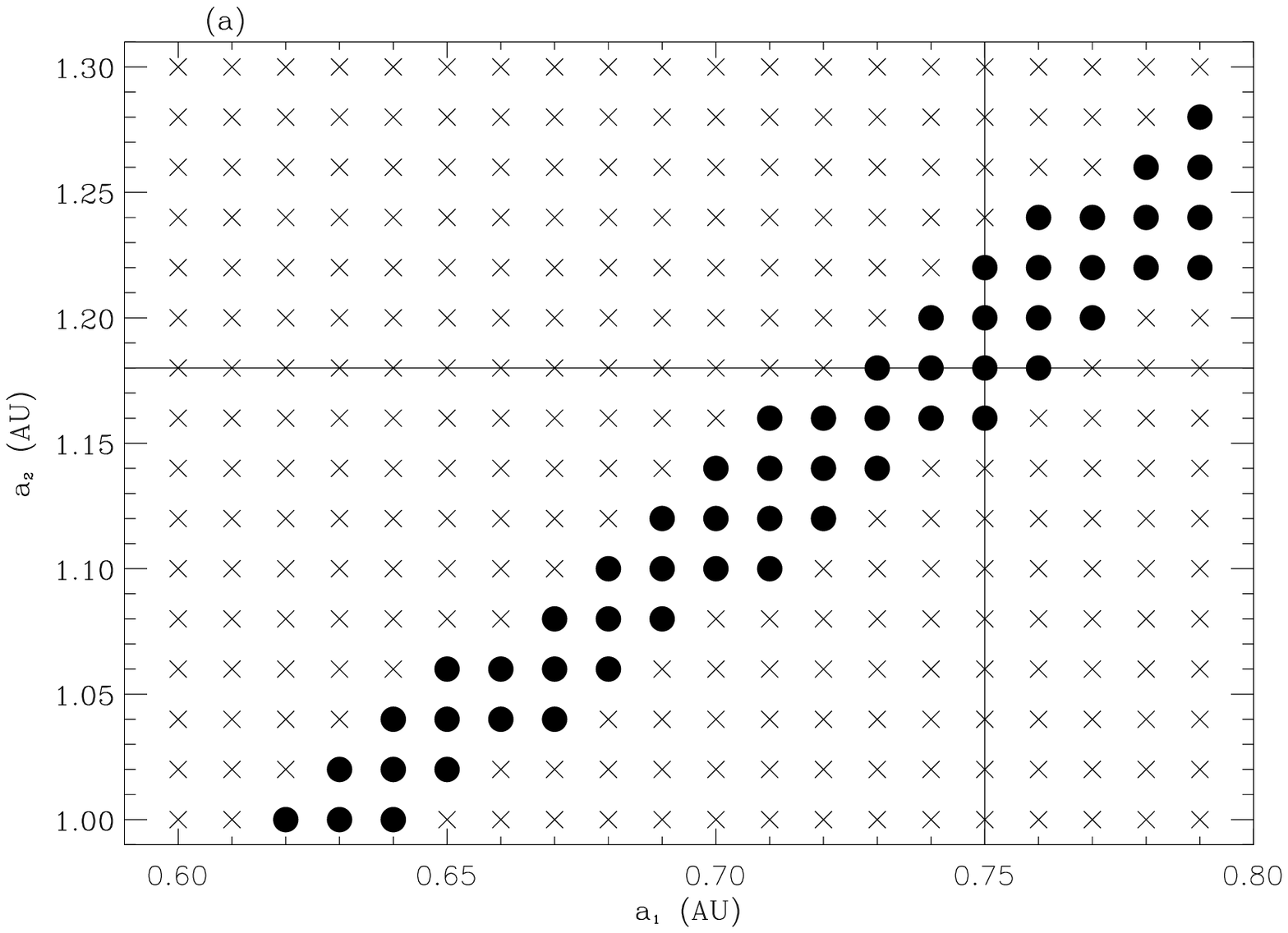}{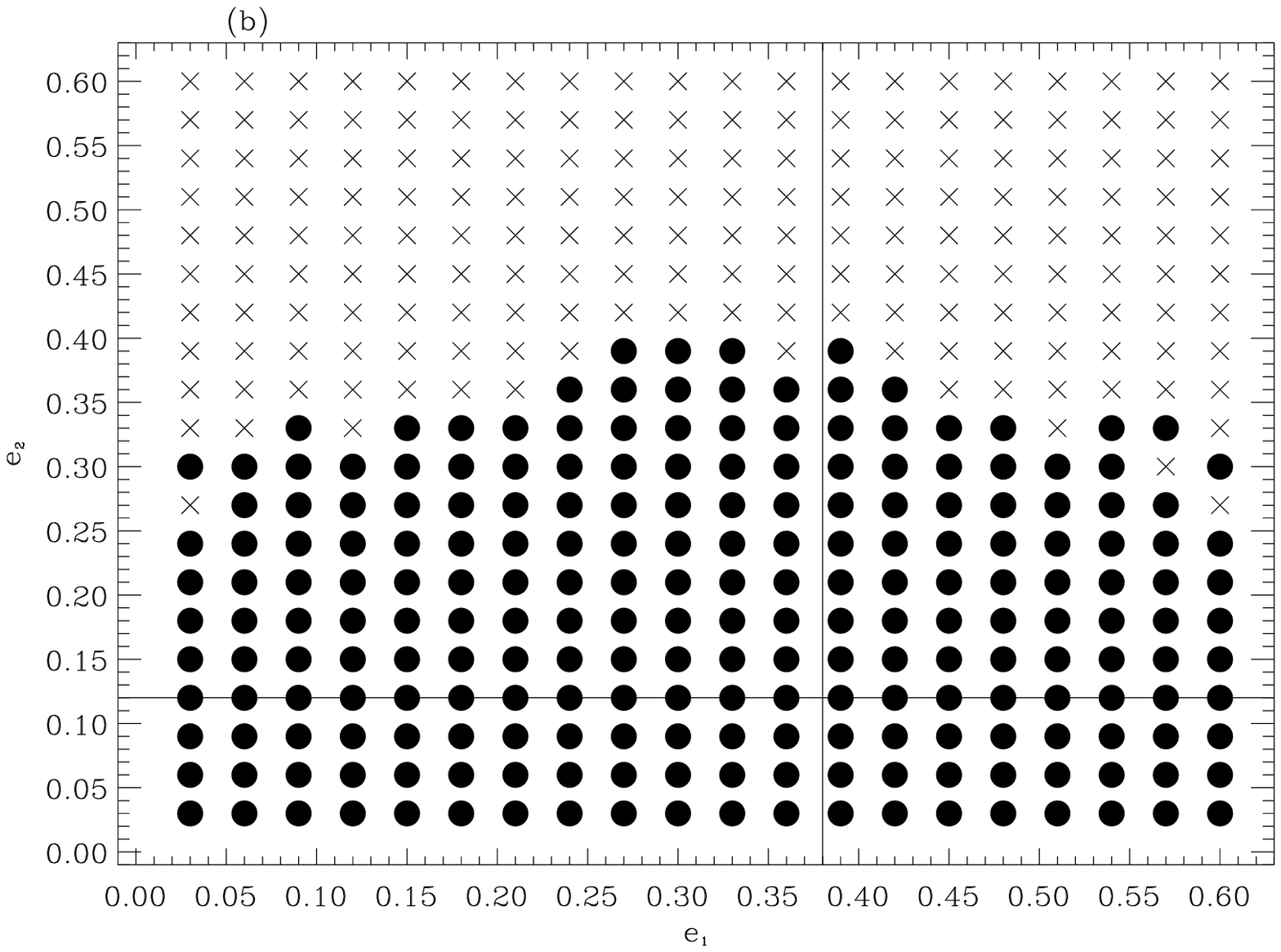} \caption{(a-b)The
stable maps in the parameter spaces of $[a_{1}, a_{2}]$ and
$[e_{1}, e_{2}]$. (c-d)The stable maps in the parameter spaces of
$[M_{1}, M_{2}]$ and $[\omega_{1}, \omega_{2}]$. The sign
\textit{crosses} indicate the unstable orbits and  \textit{filled
circles} for stable orbits lasting for $10^{6}$ yr. The
intersection between the vertical and horizontal lines represents
for the present solution from Fit 2. \label{fig7-1}}
\end{figure}

\epsscale{1.0}
\begin{figure}
\figurenum{7c-7d} \plottwo{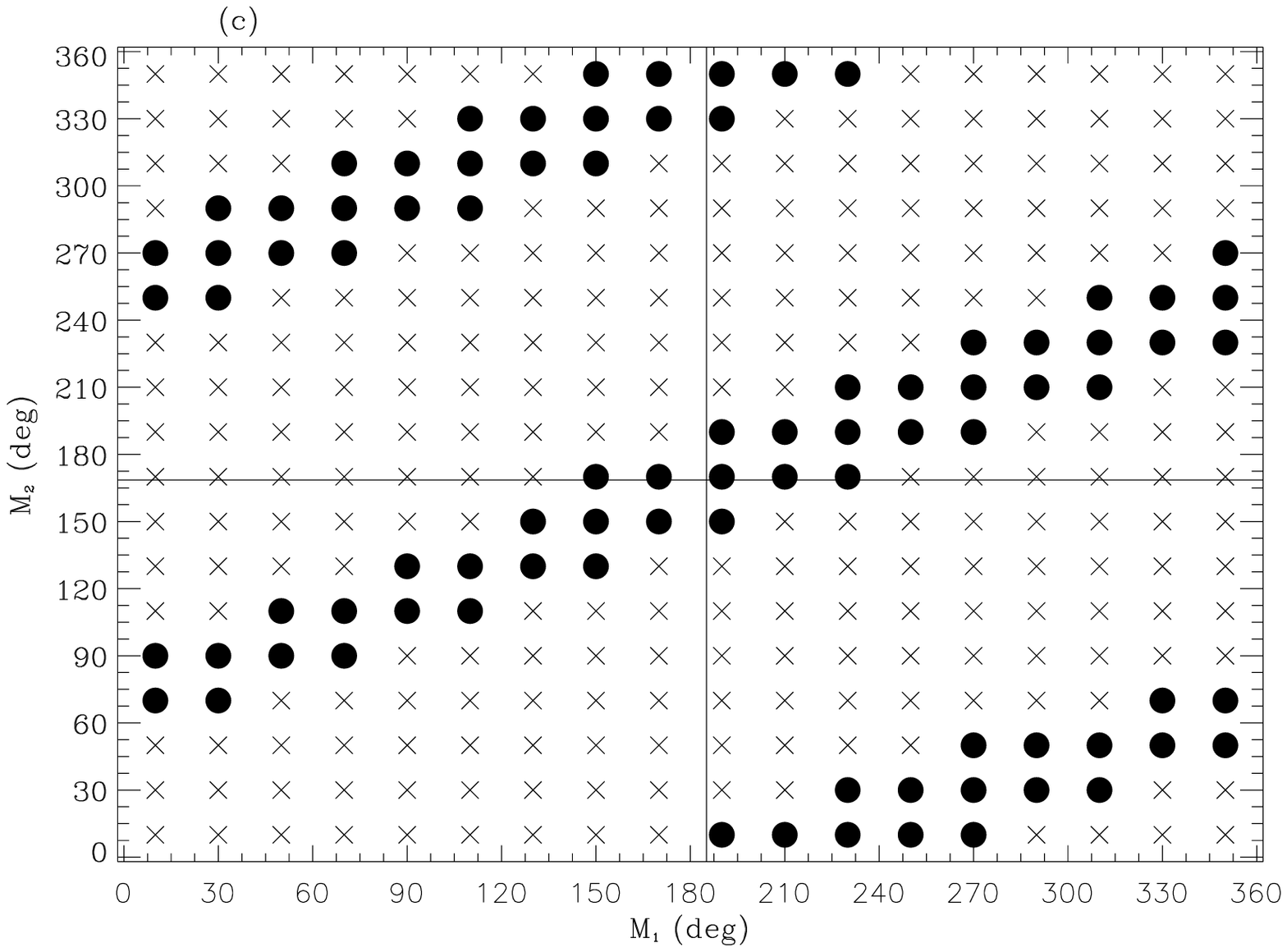}{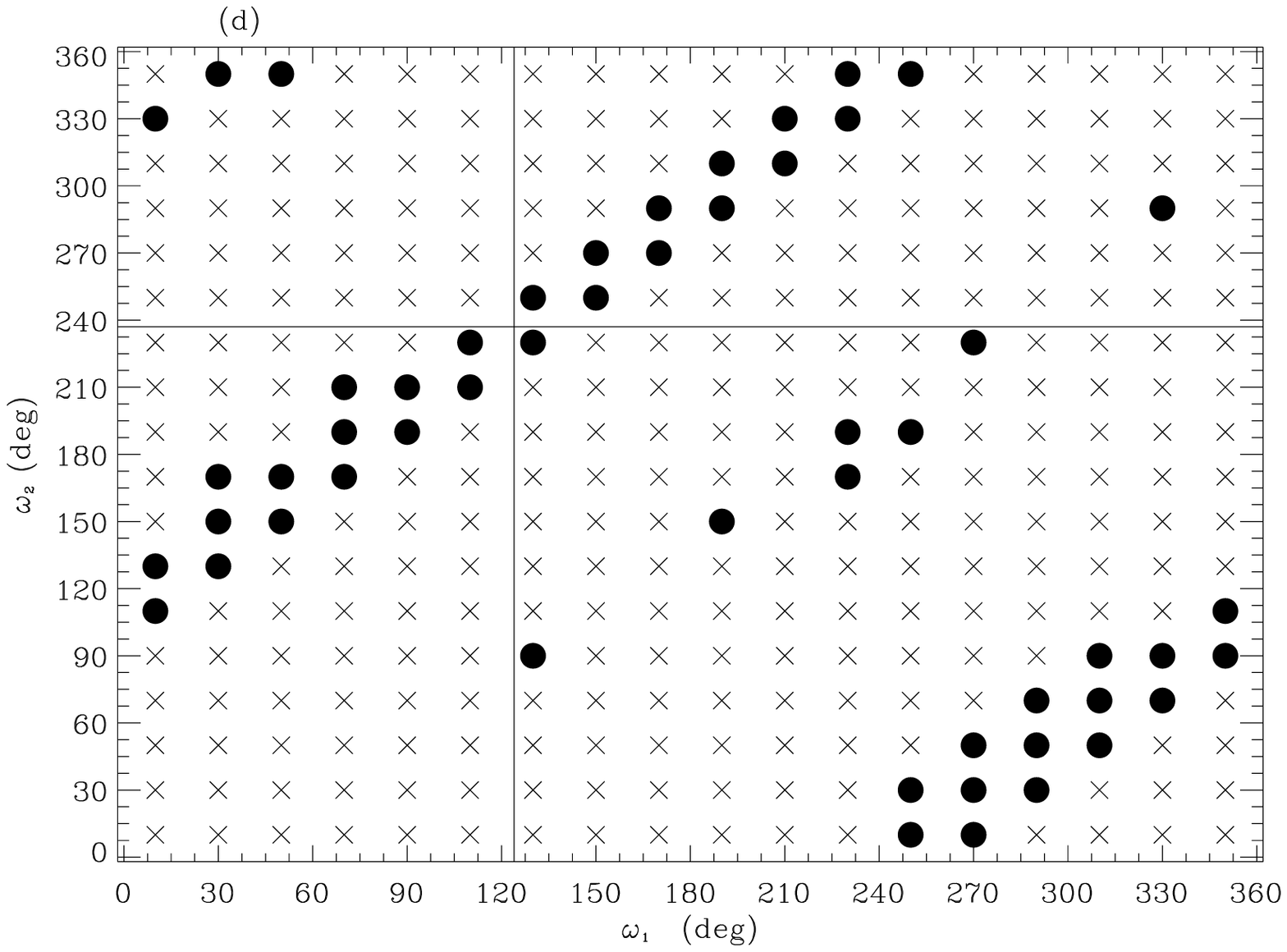}
\caption{\label{fig7-2}}
\end{figure}
\clearpage

\epsscale{1.0}
\begin{figure}
\figurenum{8} \plotone{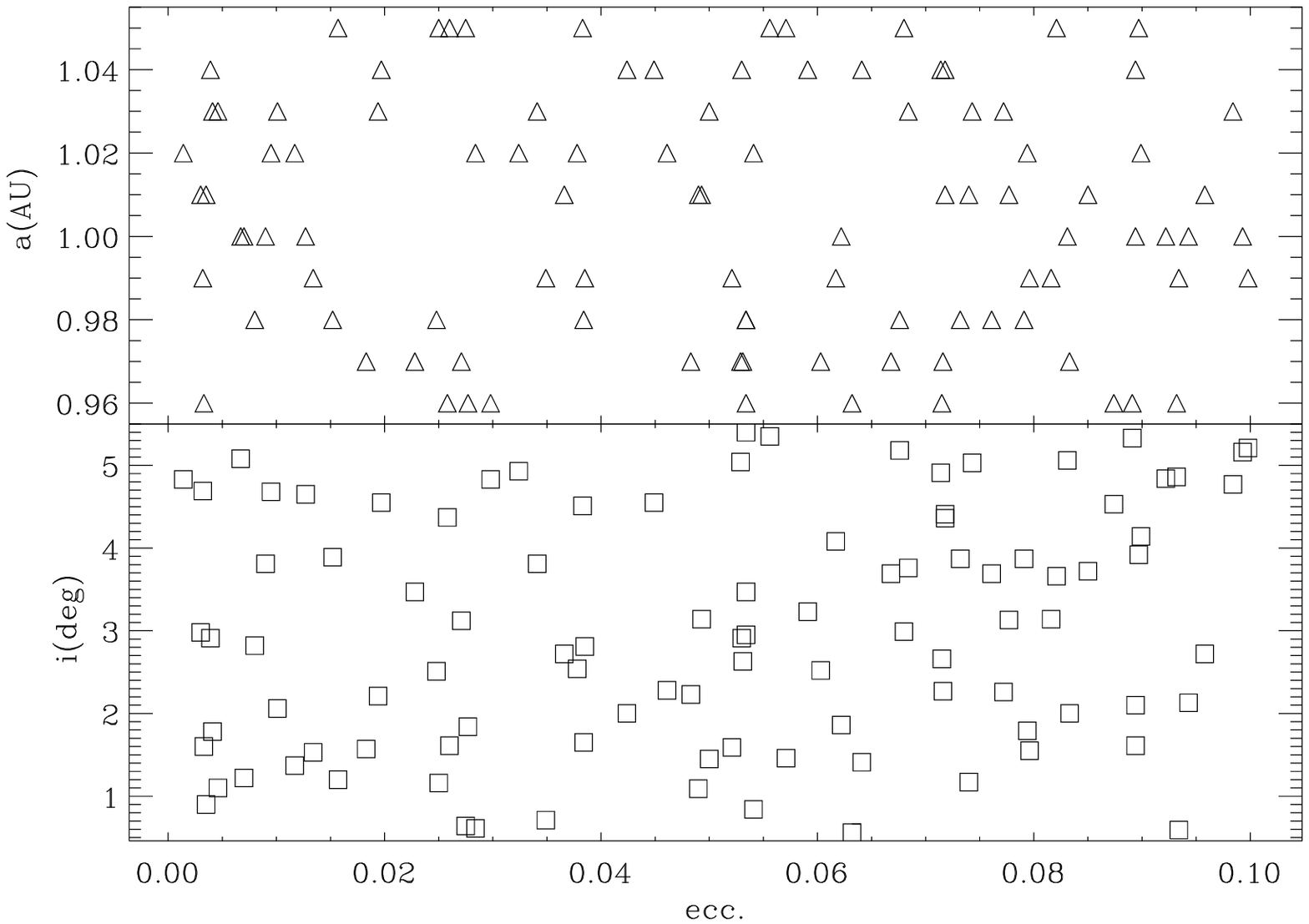} \caption{Initial orbital parameters
for 100 seed Earth-like planets. \textit{Upper panel}: the
distribution of $[e, a]$, where $a\in$ [0.96AU, 1.05AU] with the
incremental step of 0.01 AU, and $e\in [0, 0.1]$, indicating the
initial near-circular orbits for testing the dynamical
habitability. \textit{Lower panel}: eccentricity versus
inclination, where $i\in [0^{\circ}, 5^{\circ}]$, showing that
these seed planets are almost in the same plane with the resonant
pairs. \label{fig8} }
\end{figure}
\clearpage

\epsscale{1.0}
\begin{figure}
\figurenum{9} \plottwo{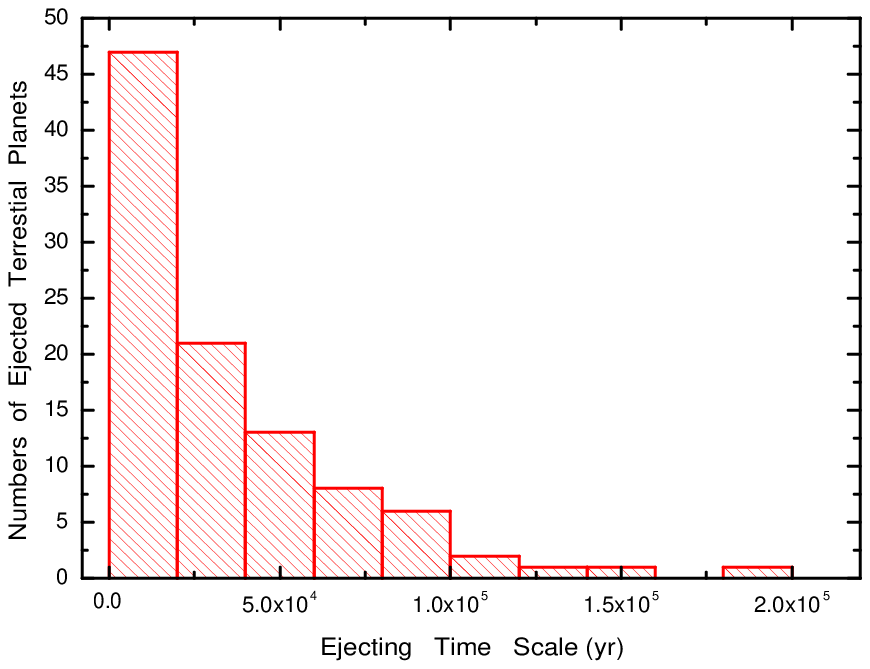}{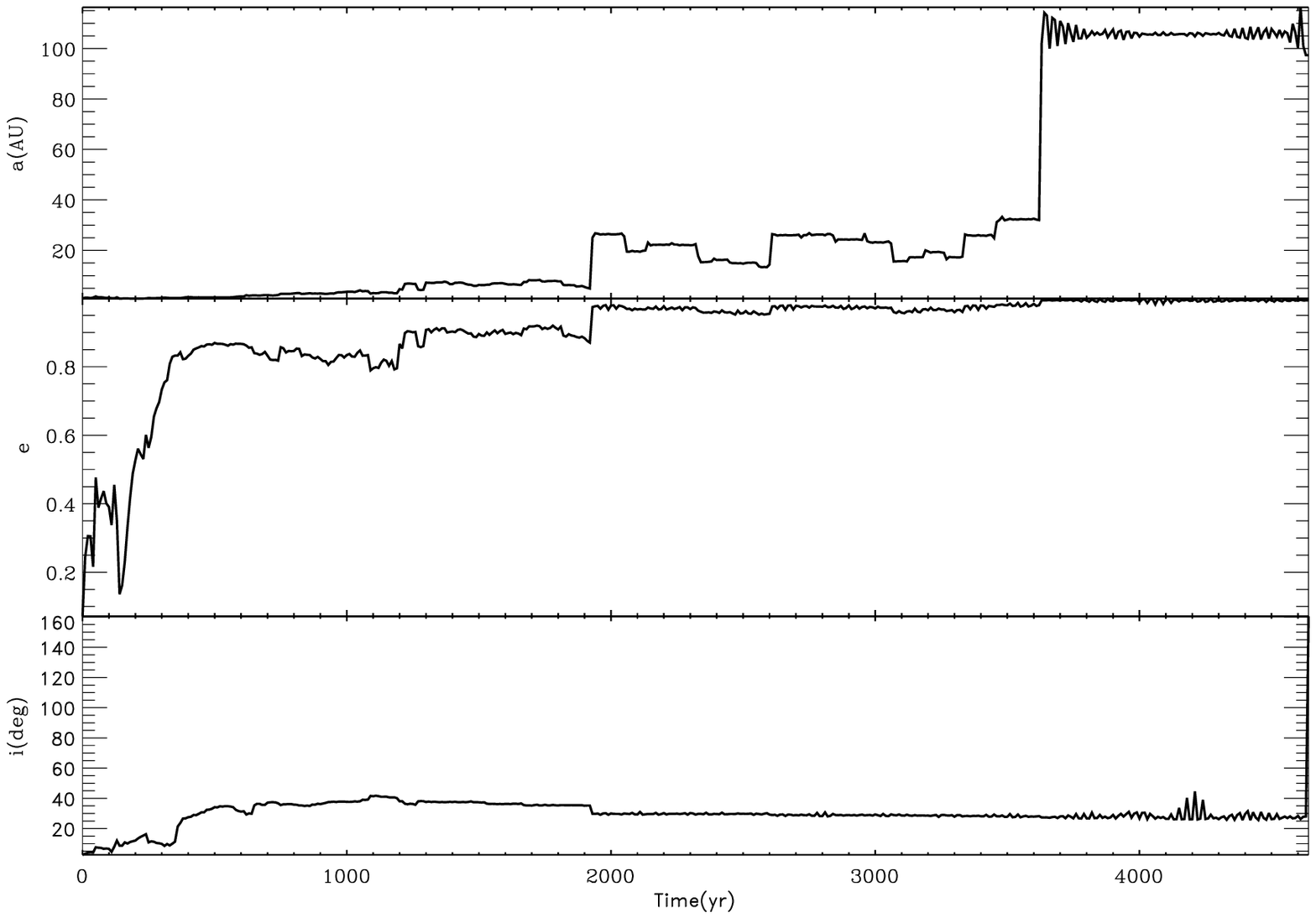} \caption{\textit{Left
panel}: (a)The numbers of the ejected terrestrial planet vs
ejecting timescale. Notice that 95\% of the orbits are ejected at
$ \tau \le 1.0 \times 10^{5}$ yr. \textit{Right panel}: (b)A
typical orbital evolution for the ejected orbits in the HD 82943
system: the semi-major axe $a$ grows from $\sim 1$ AU to over 100
AU, the eccentricity $e$ undergoes a rapid increase from $\sim
0.1$ to 1 and the inclination $i$ is also excited to a high value
of $\sim 40^{\circ}$. Finally, the assumed Earth-like planet is
ejected away from the two-planet system at $4,800$ yr.
\label{fig9} }
\end{figure}
\clearpage

\epsscale{1.0}
\begin{figure}
\figurenum{10} \plotone{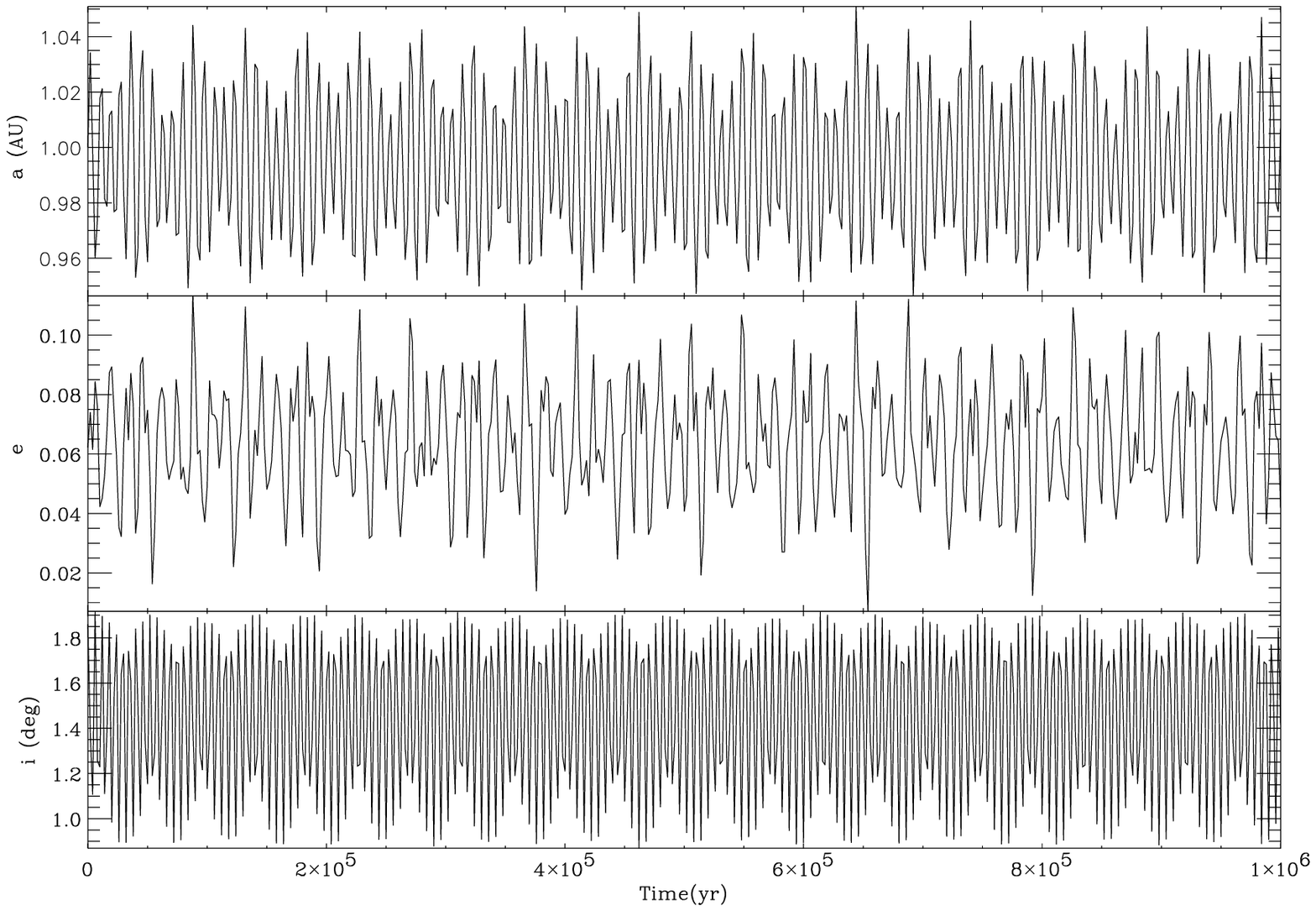} \caption{Orbital variations for
possible Earth-like planet in the GJ 876 system. Both the
semi-major axis $a$ and the eccentricity $e$ execute small
fluctuations about 1 AU and 0.06, respectively, and the
inclination $i$ also remains less than 2 degrees over the same
time span. And there is no sign to indicate that such regular
orbits at $\sim 1$ AU with lower eccentricities will become
chaotic for much longer time, even for the age of the star.
\label{fig10}}
\end{figure}

\end{document}